\documentclass[fleqn,usenatbib]{mnras}

\usepackage{newtxtext,newtxmath}
\usepackage[T1]{fontenc}

\DeclareRobustCommand{\VAN}[3]{#2}
\let\VANthebibliography\thebibliography
\def\thebibliography{\DeclareRobustCommand{\VAN}[3]{##3}\VANthebibliography}

\usepackage{graphicx}\graphicspath{{Plots/}}
\usepackage{amsmath}

\newcommand{\Like}{\mathcal{L}}
\newcommand{\Mentari}{\texttt{Mentari}}

\title[Clustering-based luminosity and mass function]{Measurement of the evolving galaxy luminosity and mass function using clustering-based redshift inference}

\author[G.\ S.\ Karademir et al.]{Geray S.\ Karademir$^1$\thanks{E-mail: gkarademir@swin.edu.au},
Edward N.\ Taylor$^1$, 
Chris Blake$^1$,
Michelle E.\ Cluver$^{1,2}$,
Thomas H.\ Jarrett$^3$, \newauthor
and Dian P.\ Triani$^{4,5}$
\vspace{6pt}\\
$^{1}$ Centre for Astrophysics \& Supercomputing, Swinburne University of Technology, PO Box 218, Hawthorn, VIC 3122, Australia; 
\\
$^{2}$ Department of Physics and Astronomy, University of the Western Cape, Robert Sobukwe Road, Bellville, South Africa; 
\\
$^{3}$ Department of Astronomy, University of Cape Town, Rondebosch, South Africa; 
\\
$^{4}$ Research School of Astronomy and Astrophysics, Australian National University, Weston Creek, ACT 2611, Australia; 
\\
$^{5}$ ARC Centre of Excellence for All Sky Astrophysics in 3 Dimensions (ASTRO 3D)
}
\date{Accepted 2023 April 24. Received 2023 March 22; in original form 2022 September 27}
\pubyear{2023}

\begin{document}
\label{firstpage}
\pagerange{\pageref{firstpage}--\pageref{lastpage}}

\maketitle

\begin{abstract}
We develop a framework for using clustering-based redshift inference (cluster-$z$) to measure the evolving galaxy luminosity function (GLF) and galaxy stellar mass function (GSMF) using WISE W1 ($3.4\mu m$) mid-infrared photometry and positions. We use multiple reference sets from the Galaxy And Mass Assembly (GAMA) survey, Sloan Digital Sky Survey (SDSS) and Baryon Oscillation Spectroscopic Survey (BOSS). Combining the resulting cluster-$z$s allows us to enlarge the study area, and by accounting for the specific properties of each reference set, making best use of each reference set to produce the best overall result. Thus we are able to measure the GLF and GSMF over $\sim 7500 \mathrm{deg}^2 $ of the Northern Galactic Cap (NGC) up to $z<0.6$. Our method can easily be adapted for new studies with fainter magnitudes, which pose difficulties for the derivation of photo-$z$s. With better statistics in future surveys this technique is a strong candidate for studies with new emerging data from, e.g. the Vera C. Rubin Observatory, the Euclid mission or the Nancy Grace Roman Space Telescope.
\end{abstract}

\begin{keywords}
galaxies:~distance and redshifts --methods: data analysis --methods: statistical
\end{keywords}
\section{Introduction}
The galaxy luminosity function (GLF) and the galaxy stellar mass function (GSMF) are two of the most fundamental descriptors in galaxy formation and evolution and are essential benchmarks for any theoretical model, counting the number of galaxies within bins of luminosity or stellar mass weighted by volume. As luminosity is only a proxy of mass, the GSMF is arguably the more astrophysically important quantity as it describes the integral over the cosmic star formation history. The general shape of the GSMF depends on internal stellar growth and mergers \citep[e.g.][]{Rodriguez-Gomez2015, Lotz2021, OLeary2021} and is similar to the shape of the GLF. While the low mass end of the GSMF is built by in-situ star formation, the high mass end is built by galaxy mergers \citep[][]{Robotham2014}. The exponential drop-off at high masses originates from a strong suppression of star formation due to AGN feedback \citep[e.g.][]{Bahe2013,Larson1980,Feldmann2015}, starvation or strangulation \citep[e.g.][]{Springel2005,Croton2006,Beckmann2017} and other effects. To test these theories, simulations are required to reproduce the observed galaxy population. Consequently, the observed GSMF and GLF are used as a benchmark for simulation studies \citep[e.g.][]{Henriques2013,Schaye2015,Pillepich2018}. Infrared emission, especially the near-infrared, is a practical tool to determine star formation rates of galaxies and is generally taken as a reasonable direct proxy for stellar mass, as it is less influenced by bright and short lived stellar populations then the optical, as well as less affected by dust attenuation \citep[e.g.][]{Bell2001,Taylor2011}.

The near-infrared GLF has been measured by \cite{Cole2001}, combining data from the 2MASS Redshift Survey \citep{Skrutskie2006} and the 2dF Galaxy Redshift Survey \cite{Madgwick2002}, deriving one of the first GLF and GSMF based on an infrared selected sample, and creating a sample free from any potential biases that affect infrared luminosity functions derived from optically-selected samples. The first total and morphologically-typed GLF measurements with statistical uncertainties comparable to the local optical GLF was measured by \cite{Kochanek2001} in the near-infrared K-band. By combining data from the 2MASS and Sloan Digital Sky Survey (SDSS) \citep[][]{York2000}, the GSMF in the local universe was measured by  \cite{Bell2003}. Multiple additional studies \cite[e.g.][]{Floc2005,PerezGonzales2005,Babbedge2006,Caputi2007,Goto2010} have followed, using spectroscopic redshifts as well as photometric redshifts. As the assembly history of galaxies depends both on galaxy type and luminosity, accurate GLF measurements as a function of redshift are needed. \cite{Dai2009} measured the GLF for early- and late-type galaxies up to $z<0.6$ using IRAC data from the Spitzer Space Telescope \citep[][]{Werner2004} in the near- and mid-infrared. It was found that the GLF of late-type galaxies resembles that of the total population. Whilst it has similar characteristic magnitudes, the GLF of early-type galaxies shows deviations from flat luminosity density evolution.

Using dominantly photometric redshifts, several studies have measured the infrared GLF within $0<z<3$ \citep[e.g.][]{Muzzin2013,Davidzon2017,Wright2017,Adams2021} and up to $z\sim7$ \citep[e.g.][]{McLure2009,Grazian2015,Furtak2021}. 
Despite the large efforts undertaken, there is no agreement on the exact shape of the GSMF. Studies of the high redshift GSMF, have for example found much steeper slopes at the low mass end compared to studies at lower redshifts \citep[e.g.][]{Song2016}. Moreover studies which find similar constant characteristic mass with redshift disagree by up to $M_* \sim 0.5$ dex \citep[e.g.,][]{McLeod2021, Thorne2021}. The presence of cosmic variance within and between surveys leads to statistical and systematic uncertainties, and the usage of photometric redshifts in particular can lead to systematic uncertainties in the shape and evolution of the luminosity functions.

One of the most recent and all-sky surveys in the near- to mid-infrared is the Wide-field Infrared Survey Explorer (WISE) \citep[][]{Wright2010}. While having measured millions of sources, WISE is particularly suited for the task of focusing on the stellar mass, as the $3.4 \mu m$ (W1) and $4.6 \mu m$ (W2) bands are tracing the continuum emission of low-mass, evolved stars with little sensitivity to the interstellar medium (ISM) through emission or absorption \citep[e.g.][]{Jarrett2011, Cluver2014}. This makes W1 the ideal waveband to study the stellar masses of galaxies.

Exploiting WISE for galaxy evolution has been limited by the availability of redshifts. For example, \cite{Jarrett2017} has established an $z \lesssim 0.5$ sample with carefully deblended photometry measuring the stellar mass distribution with modelled source counts up to $z\sim2$ within $60 \, \mathrm{deg}^2$ by cross-matching WISE with spec-$z$s from the Galaxy And Mass Assembly (GAMA) survey. \cite{Donoso2012} cross-matched the WISE data with SDSS and investigated $\sim 95{,}000$ galaxies to derive the $z\sim0.1$ galaxy luminosity function of WISE galaxies, $28\% $ of W1 sources have faint or no r-band counterparts ($m_r > 22.2$) with SDSS \citep{Yan2013}.

Our goal is to unlock the full information potential of WISE through the use of clustering-based redshift inference (cluster-$z$s), measuring the GLF and GSMF. This technique makes use of the fact, that galaxies are not homogeneously distributed, but instead cluster strongly, both in real space and in projection on the sky \citep[e.g.][]{Peebles1980, McNaught2014, Jarrett2017}. The idea of using the angular clustering of galaxies to infer their redshift distribution has been discussed for several decades \citep[e.g.][]{Seldner1979, Phillipps1985, Phillipps1987}, but has only started to get used recently. This technique has been investigated by \cite{Schneider2006} measuring cross-correlations of galaxies binned by photometric redshift. The formal approach of computing redshift distributions by measuring the angular cross correlation between a photometric sample and different redshift bins of a spectroscopic sample has been outlined by \cite{Newman2008} and \cite{Matthews2010, Matthews2012}. Techniques measuring the cross-correlation over fixed physical scales have been proposed by \cite{Schmidt2013} and \cite{Menard2013}. Recent studies applying this technique to derive redshift distributions include, e.g. \cite{McQuinn2013, Rahman2015, Rahman2016a} and \cite{Scottez2016}. \cite{Menard2013} derived the redshift distribution for three different colour samples of WISE, showing that the technique is not limited to the optical. Clustering-based redshifts hence provide valuable information on the redshift dimension of astronomical datasets and could be used as a primary redshift estimator by the next generation of cosmological surveys \citep[][]{Scottez2018}. Instead of just inferring the redshift distribution, \cite{vanDaalen2018} showed how clustering-based redshift inference can be used to measure the galaxy luminosity function for simulated data. \cite{Bates2019} demonstrated how the stellar mass function and i-band luminosity functions for SDSS galaxies can be derived using cluster-$z$s and in \cite{Karademir2022} it was shown how cluster-$z$s can be used to probe the r-band GLF more than a decade in luminosity beyond the limits of the GAMA spectroscopic redshift limit, down to where Globular Clusters take over as the most numerous extra-galactic population.

This paper is structured as follows: In Sec. \ref{sec:Data} we describe the different datasets used in this study: the imaging and photometry datasets as well as the reference datasets that are used to calculate the cluster-$z$s. The methodology for calculating the cluster-$z$s is described in Sec. \ref{sec:cluster-$z$s}. The parametric description of the model in addition to the fitting process is shown in Sec. \ref{sec:model} and Sec. \ref{sec:mcmc}. We show the resulting GLF and GSMF in Sec. \ref{sec:results}, which we discuss and summarize in Sec. \ref{sec:discussion} and Sec. \ref{sec:summary}. Throughout the paper we use a flat $\Lambda$CDM cosmology with $\Omega_M=0.3$, $\Omega_\Lambda=0.7$ and a Hubble parameter $H_0 = 100 \ h$ km Mpc$^{-1}$ s$^{-1}$ where $h = 0.7$.

\section{Data}
\label{sec:Data}
In this section, we describe the three distinct components of our analysis from which we will derive the evolving WISE GLF and GSMF. First, in Sec. \ref{sec:target-data} we describe the WISE photometric catalogue that is the target of our cluster-$z$ analysis. Then, in Sec. \ref{sec:ref-data}, we describe the spectroscopic redshift catalogues that we used as our reference datasets to infer the redshift distribution of WISE sources based on cross-correlation clustering statistics. As we discuss in Sec. \ref{sec:cluster-$z$s}, our cluster-$z$ results are degenerate with the galaxy bias of the target sample and only proportional to the redshift distribution up to an unknown scalar. To properly normalize the galaxy density, we require an external constraint. The $z \sim 0$ GAMA dataset we use to anchor our GLF and GSMF results is described in Sec. \ref{sec:specz}.

\subsection{WISE Photometric Data}
\label{sec:target-data}
The Wide-Field Infrared Survey Explorer (WISE) \citep[][]{Wright2010} is a full sky, near- to mid-infrared survey in four different wavebands centered on $3.4 \mathrm{\mu m}$ (W1), $4.6 \mathrm{\mu m}$ (W2), $12 \mathrm{\mu m}$ (W3), and $22 \mathrm{\mu m}$ (W4). The telescope has an $40$cm aperture providing a field of view of $47\arcmin\times47\arcmin$ at an angular resolution of $6.1\arcsec$, $6.4\arcsec$, $6.5\arcsec$, and $12.0\arcsec$ at the respective wavebands. Starting operation on 14 January 2010, full sky completion was achieved on 17 July 2010. After the secondary solid hydrogen cryogen tank was depleted on 5 August 2010 observations in W4 were terminated, while observations in the remaining three bands continued until the primary tank was exhausted on 29 September 2010. With the NEOWISE Post-Cryogenic Mission \citep{Mainzer2011} observations within W1 and W2 bands continued until 1 February 2011. After the spacecraft was put into hibernation, NEOWISE has been reactivated with the task of discovering and characterizing near-Earth objects (NEO) and the first data were obtained on 7 December 2013 \citep{Mainzer2014}. While operations were expected to end in 2017, NASA decided to extend the NEOWISE mission until at least June 2023. With the AllWISE Data Release, based on the AllSky public release data, the source catalogue extended to nearly 750 million sources \citep[][]{Cutri2013}. These data however were intentionally blurred by convolution of their point-spread function, since the AllWISE catalogue is optimized for point sources. Considering the large number of sources, this results in overlap and therefore loss of many sources.

\begin{figure}
	\centering
	\includegraphics[width=\columnwidth]{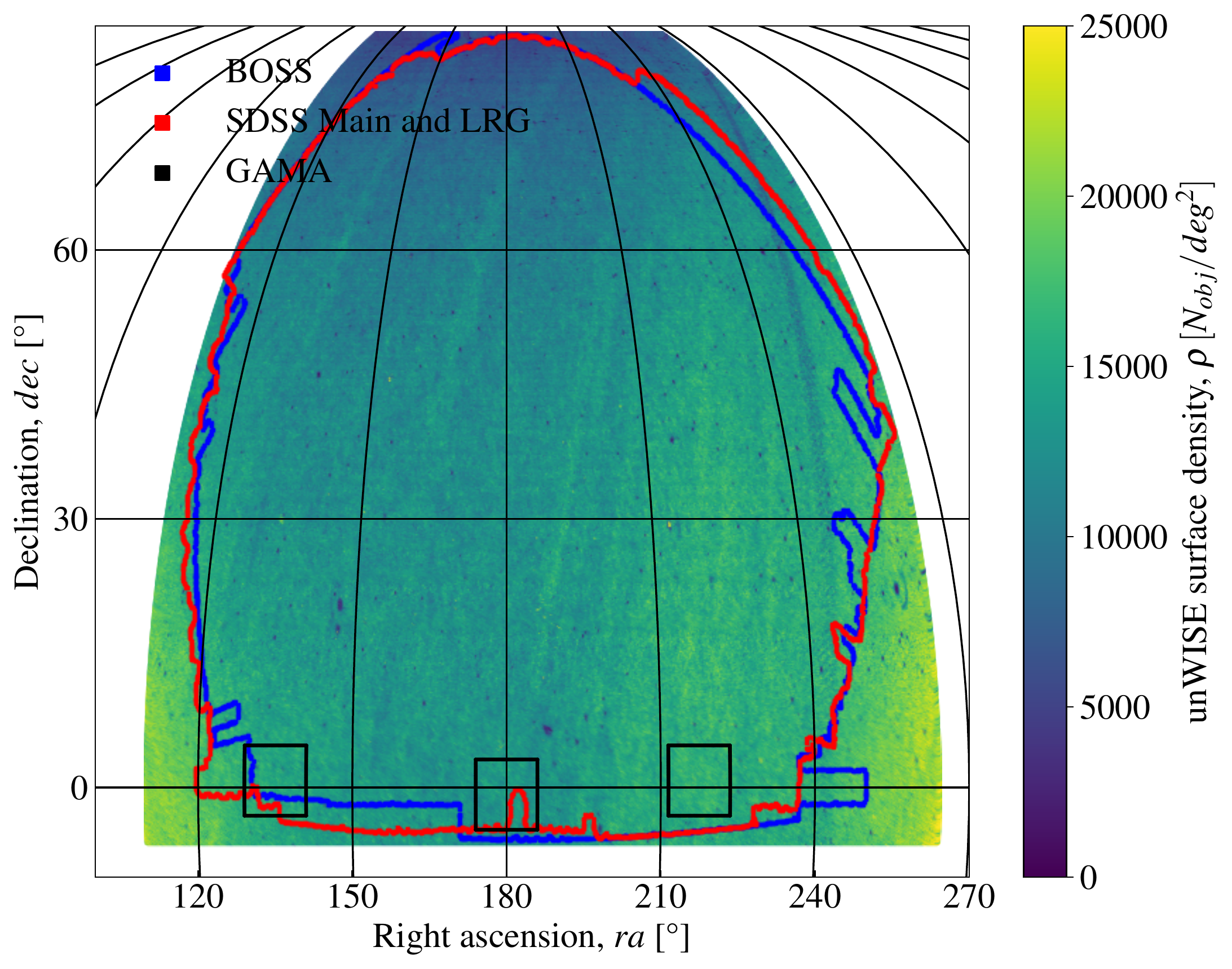}
	\caption{The unWISE surface density over the full study area. The edges of the study for the different reference sets are shown on top of the unWISE data in blue for BOSS, red for SDSS Main and LRG as well as in black for GAMA. unWISE provides an approximately spatially uniform distribution within the study area.}
 	\label{fig:Moll}
\end{figure}

With the analysis of unblurred coadds of the WISE and NEOWISE images, the resulting unWISE catalogue \citep[][]{Lang2014, Meisner2017, Meisner2017b} is able to reach an additional $\sim0.7$ magnitudes in-depth, resulting in the detection of roughly three times as many sources (about two billion over the entire sky for W1 and W2) compared to AllWISE. Due to the relatively large full width at half maximum (FWHM) of the point spread function (PSF) of $\sim 6 \arcsec$ at the W1- and W2-band, all sources are treated as unresolved and are approximated as point sources \citep[][]{Schlafly2019}. The fluxes are derived from the best fit PSF-model using the \texttt{crowdsource} analysis pipeline \citep[][]{crowdsource}. The resulting unWISE dataset provides a spatially uniform target dataset for our study (see Fig. \ref{fig:Moll}).

In this study, we only use the W1 flux at $3.4 \mu m$ up to a completeness limit for W1 at $m_{W1,\mathrm{Vega}} = 17.5$ derived by \cite{Jarrett2017}, which is slightly brighter than the confusion limit of W1 at $m_{W1,\mathrm{Vega}} = 18.1$ \citep{Jarrett2011}. unWISE provides us with a total of $\sim 2.7\times 10^8$ objects within our completeness limit within the coverage of the SDSS and BOSS surveys in the northern galactic cap (see details about the area selection for each reference set in the next section). This results in an average object density of $\sim 3.6 \times 10^{4}\, \mathrm{per \, deg}^2$. We convert W1 Vega Magnitudes to AB magnitudes assuming $m_{W1,\rm AB} - m_{W1,\rm Vega} = 2.699$ \citep{Jarrett2011}. All magnitudes mentioned below are within the AB magnitude system, unless otherwise stated. 

\subsection{Spectroscopic Redshift Reference Sets}
\label{sec:ref-data}
As described in Sec. \ref{sec:cluster-$z$s}, clustering redshift inference works by considering angular cross-correlation statistics between the positions of the target dataset with the positions for a reference dataset with known redshifts. The function of the reference dataset is to trace the large-scale cosmic structure (LSS) and to project the target dataset with it. As such, the analysis is limited by how well the reference set maps cosmic structure in the field. The ideal reference dataset would span the widest possible area, with the highest possible source density, and covers the largest possible redshift interval. Unfortunately, there isn't the "one" dataset which satisfies all these requirements and in practice, the choice of reference set involves a trade-off between these different factors. As we describe below, we use several different spectroscopic redshift datasets to provide a robust measure of the WISE source redshift distribution for $z < 0.6$.

The largest area redshift surveys to date have been accomplished by the SDSS and its successors. These surveys cover thousands of square degrees on the North Galactic Cap (NGC) and are therefore the preferred first choice for our reference set.

For the low to mid redshift range, we use data from the SDSS Legacy survey \citep[][]{York2000, Stoughton2002, Gunn2006}. We use data from DR16 \citep[][]{SDSSDR162020} selecting galaxies of the main galaxy sample \citep[][]{Strauss2002} as well as the luminous red galaxy (LRG) sample \citep[][]{Eisenstein2001}. We use all the data within the NGC, excluding the strip northeast of the main footprint to create a contiguous study area. This results in a total area of $\sim 7200 \mathrm{deg}^2$ and $\sim1.7\times10^5$ galaxies for the LRG and $\sim6.4\times10^5$ for the main survey.

The SDSS cluster-$z$s are calculated using randoms from the NYU Value-Added Galaxy Catalogue (NYU-VAGC) \citep{Blanton2005,Adelman2008,Padmanabhan2008}. The NYU-VAGC is a cross-matched collection of galaxy catalogues including carefully constructed large-scale structure samples. This catalogue has been created for the study of galaxy formation and evolution. We use the random data from the large-scale structure samples of DR7. For these catalogues, the target, and tiling masks are combined for the imaging and the flux limit and completeness are tracked as a function of position

For the higher-$z$ coverage we use data from the SDSS-III:BOSS \citep[][]{Eisenstein2011, Dawson2013, Smee2013} survey. The Baryon Oscillation Spectroscopic Survey (BOSS) is a cosmological survey that measured spectroscopic redshifts for $~1.5$ million luminous red galaxies between $0.2 \lesssim z \lesssim 0.8$ covering roughly $10,000 \mathrm{deg}^2$ on the sky. It aimed to measure the scale of the baryon acoustic oscillations (BAO) in the clustering of galaxies. In our study, we use BOSS DR12 data \citep[][]{Alam2015} from the NGC resulting in an effective area of $\sim7500 \, \mathrm{deg}^2$ and $\sim 9.5\times10^5$ objects. To correct for selection effects we use the galaxy weights: $w_{\mathrm{tot}} = w_{\mathrm{systot}}(w_{\mathrm{cp}}+w_{\mathrm{noz}}-1)$ \citep[see][]{Reid2016}, but no FKP weights. For the random dataset for BOSS, we use the pre-generated random catalogue from the BOSS collaboration.

The third reference dataset we use is the Galaxy And Mass Assembly (GAMA) redshift survey \citep[][]{Driver2022}. GAMA provides data with a $> 99$\% redshift completeness to the $m_r < 19.8$ selection limit over the three equatorial regions G09, G12 and G15. It has no discernible incompleteness as a function of pair separation \citep{Liske2015} and a much higher source density than SDSS or BOSS. While covering only $180 \, \mathrm{deg}^2$ it provides us with $\sim 1.9\times10^5$ spec-$z$ measurements. For GAMA we use the randoms with the same selection function as the real galaxies provided by the GAMA collaboration \citep[][]{Farrow2015}. 

\begin{figure}
	\centering
	\includegraphics[width=\columnwidth]{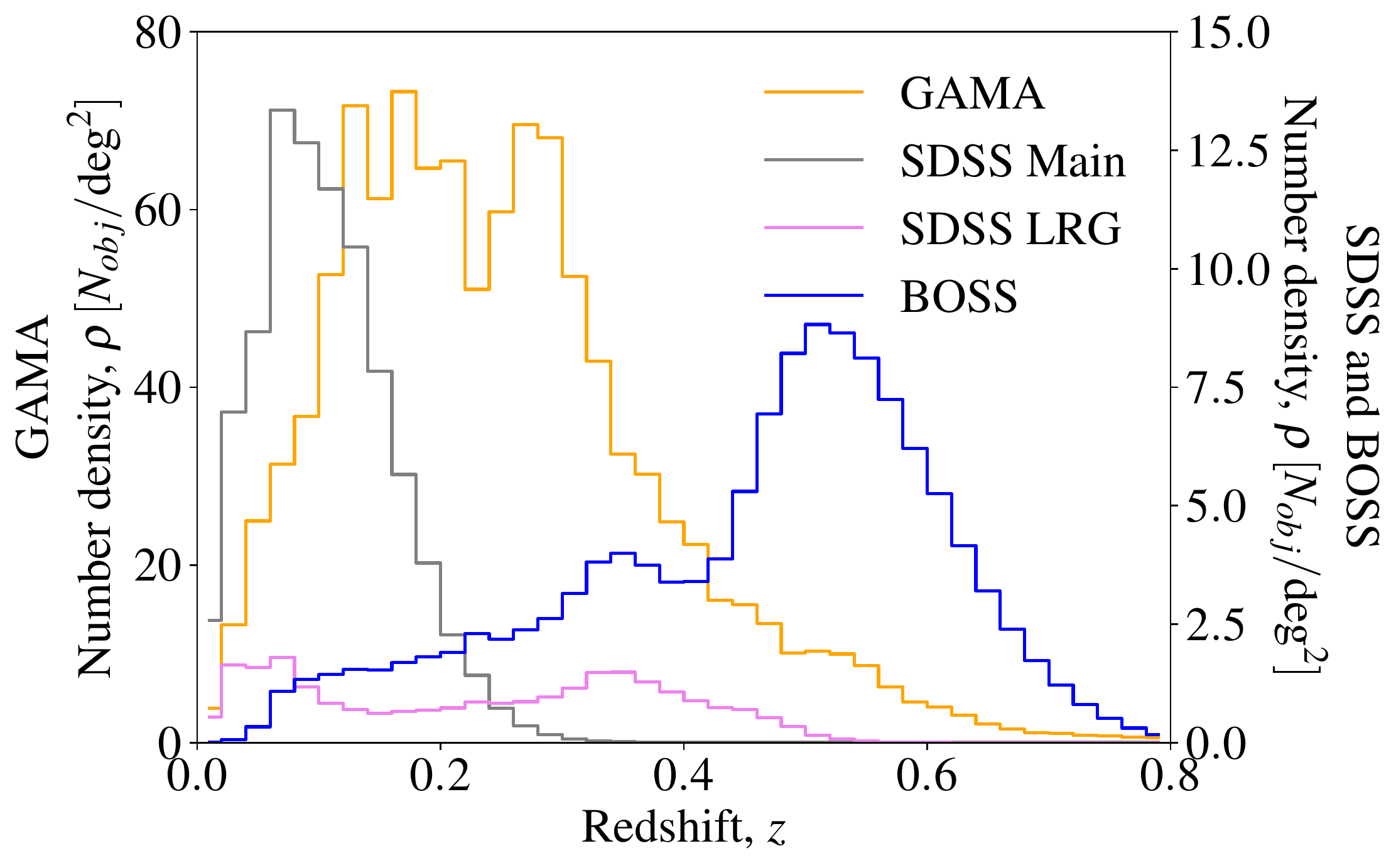}
	\caption{Density distribution of the SDSS, BOSS and GAMA reference datasets per square degree vs redshift. The redshift ranges of each reference set build up the extended full redshift range probed.}
 	\label{fig:redshift_distri}
\end{figure}

In combination these datasets provide a reference set, covering a large area and providing large numbers of redshift measurements as well as a sufficiently large redshift baseline up to $z<0.8$. The exact ranges used for the measurement of the GLF are shown in Sec. \ref{sec:Nz}.

\subsection{The spectroscopic anchor dataset}
\label{sec:specz}
For the determination of the GLF, we need an external constraint on the characteristic galaxy density $\phi^*_0$ (see Sec. \ref{sec:mcmc} for further details) at $z\sim0$. For this task, we use data from the Galaxy And Mass Assembly (GAMA) survey DR4 \citep[][]{Driver2022} as it consists of spectroscopic redshifts for a W1 flux-limited sample.

\begin{figure}
    \centering
    \includegraphics[width=\columnwidth]{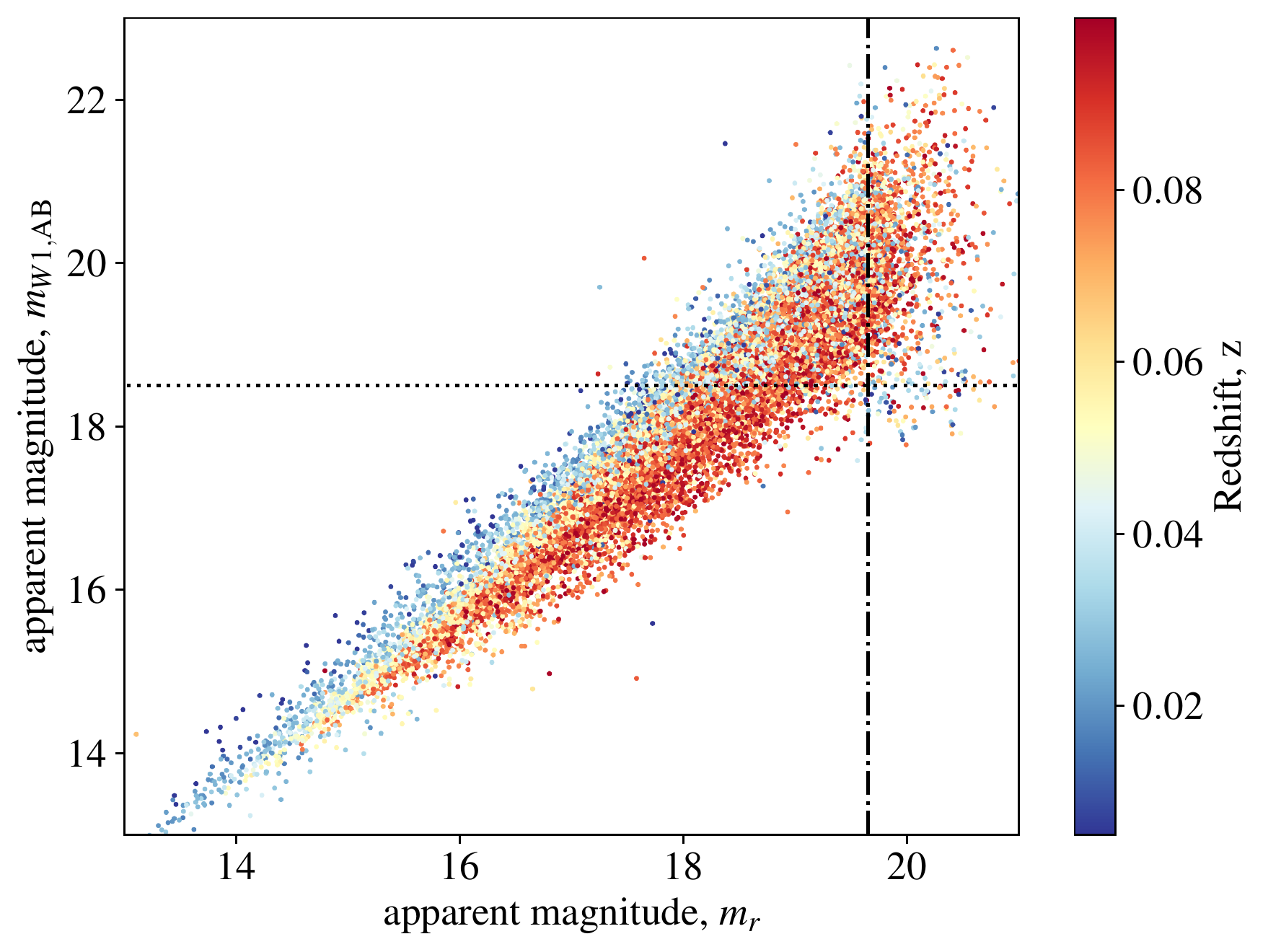}
    \caption{The $\mathrm{mag}_{W1}$ vs $\mathrm{mag}_r$ distribution of GAMA galaxies at $z<0.1$ within the equatorial regions. The scatter points are colored according to their redshift. The vertical dotted line indicates the GAMA completeness limit at $\mathrm{mag}_r = 19.65$, while the horizontal dotted line shows the completeness limit for $\mathrm{mag}_{W1}=18.5$ within GAMA.}
    \label{fig:GAMA}
\end{figure}

We use the three equatorial regions of GAMA (G09, G12 and G15), which cover $180 \, \mathrm{deg}^2$. While WISE photometry is optimized for point sources, significant effort has been undertaken to reconstruct the image mosaics to extract and measure the flux of extended sources \citep{Cluver2014} using isophotal apertures for extended resolved sources. The resulting WISECat dataset reaches a $10\sigma$ depth of $m_{W1,\rm Vega} = 16.6$ \citep[see ][]{Cluver2020}. From the study of the WISECat photometry, it was seen that a profile fitting approach, as used by unWISE, is missing flux, especially at low z, with a mean difference of $\sim0.2$mag compared to aperture fitting measurements of the same objects (see Appendix for further details). To obtain the initial galaxy density at low redshift we only select objects at $z<0.1$ and $m_{W1,\rm AB}<18.5$ from WISECat (see Fig. \ref{fig:GAMA}) using the $8.25\arcsec$ standard aperture from WISE. As the GAMA spec-$z$s are complete at this redshift, we obtain a W1 limited sample of $10{,}060$ spec-$z$ measurements for our analysis.

Remembering that the main use of these data is to normalise our GLF and GSMF results by constraining the value of the characteristic density, $\phi_0$, it is significant that the GAMA regions contain cosmic sampling variance of $\pm 25\%$ for each individual $60 \, \mathrm{deg}^2$ region \citep{Driver2010}. In addition, the total $180 \, \mathrm{deg}^2$ GAMA survey is under-dense by $\sim 12\%$ relatively to SDSS. We account for this under-density by applying a redshift zero correction of $0.0866$ to the $\log(\phi^*_0)$ value of the model \citep[see][]{Driver2022} in the fitting process.

\section{Methodology}
\label{sec:methodology}
In this study we follow a similar approach to that described in \cite{Karademir2022}. The first step is the calculation of the cluster-$z$s (Sec. \ref{sec:cluster-$z$s}) for the target data (Sec. \ref{sec:target-data}) using each reference dataset described in Sec. \ref{sec:Data}. Secondly, we perform a single Schechter fit (Sec. \ref{sec:model}) using the resulting cluster-$z$s and the spectroscopic dataset from Sec. \ref{sec:specz}. After obtaining the best fitting model and the normalisation parameters $A_m$ we obtain the GLF. The resulting observers' frame results are corrected by global k-corrections described in Sec. \ref{sec:kcorr} and finally the resulting, k-corrected, GLF is transformed into the GSMF using the mass-to-light ratios described in Sec. \ref{sec:MLratio}.

\subsection{Clustering based redshift inference}
\label{sec:cluster-$z$s}
Using clustering-based redshift inference (cluster-$z$s) we can derive the redshift distribution for a set of target objects statistically, only using their positional information. In this method the positions of a target sample on the sky are cross-correlated with the positions of a reference sample, for which the redshifts are known. By comparing the amplitude of the resulting 2D angular cross-correlation over several redshift bins, the redshift distribution is obtained. The main uncertainty of this method is the degeneracy of the resulting redshift distribution with the galaxy bias of the target sample \citep[][]{Gatti2018}. This technique allows for the determination of the redshift distribution for an ensemble of objects, but does not provide redshifts for individual galaxies. The technique of deriving clustering-based redshifts is described in detail by \cite{Schneider2006} and \cite{Menard2013}. 

For the calculation, there are three samples needed. Firstly, a dataset consisting of objects with unknown redshifts but known spatial positions ($ra$, $dec$). In this study, this data is from the unWISE data described in Sec. \ref{sec:target-data}. The second dataset is a reference set with objects of accurate measurements of their full 3D positions. This dataset traces the large-scale structure and we use data from the GAMA, SDSS Main, SDSS LRG and BOSS survey (see Sec. \ref{sec:ref-data}) to create four different sets of cluster-$z$s. It is not necessary that the target and reference set consist of objects of the same colour, morphology or type, but they have to cover the same area of the sky as the target dataset. We use a random set corresponding to each reference catalogue to perform the clustering measurement.

The basic idea behind the calculation of cluster-$z$s is that the spatial distribution of galaxies on the sky is unique at each redshift. This leads to the principle that if two populations of objects overlap on the sky, but are at different redshifts, their angular correlation is expected to be zero (ignoring gravitational lensing effects).

As explained in greater detail by \cite{Menard2013}: in an ideal case where a target sample of $N_t$ galaxies is located at a narrow redshift range $z_t$, it is possible to measure the redshift distribution by splitting the reference sample into redshift bins $z_i$ and calculating the angular cross-correlation $w_{tr}(z_i)$ for each bin. Here only at $z_t$, a signal is received and the redshift distribution of the target sample is:

\begin{align}
\frac{dN}{dz} = N_t\delta(z_t-z_i) \propto w_{tr}(z_t)
\end{align}

In the typical case where the target sample is extended in redshift, the clustering amplitude can be related to the underlying clustering bias and redshift distribution as:

\begin{align}
\bar{w}_{tr}(z) \propto \frac{d P}{d z}(z) \bar{b}_{t}(z) \bar{b}_{r}(z) \bar{w}_{DM}(z) ~ .
\label{eq:w_tr}
\end{align}

The spatial cross-correlation of the target data with the reference data is the product of the shape of the redshift probability distribution $\frac{d P}{d z}(z)$, the galaxy bias factors of the reference and the target sample, respectively $\bar{b}_{r}(z)$ and $\bar{b}_{t}(z)$, as well as the dark matter clustering amplitude $\bar{w}_{DM}(z)$. Here the angular correlation is measured over a fixed physical range $r_c$, which is represented by the bar above the related quantities.

The optimal integration limits of $r_c$ depend on the particular target and reference samples as well as the validity of the assumption of linear bias over the physical scales probed. In our analysis, the lower limit has to be larger than the fibre collision radius as the angular correlation functions deviate at smaller ranges \citep[][]{Gordon2018}. The fibre collision radius for our reference sets are $55\arcsec$ for SDSS and $62\arcsec $ for BOSS \citep[][]{Dawson2013}. The upper limit has to been chosen such as it is large enough to capture the LSS and not too large so that the signal is reduced by statistical noise due to uncorrelated background galaxies. We tested multiple upper limits up to $10 \, \mathrm{Mpc}$ for the clustering ranges and it was seen that the results are stable to changes of the upper limit within these ranges. For our study we have chosen to use clustering ranges of $0.4 \, \mathrm{Mpc} < r_c < 5 \, \mathrm{Mpc}$ for SDSS, $0.8 \, \mathrm{Mpc} < r_c < 5 \, \mathrm{Mpc}$ for BOSS and for GAMA we use smaller ranges of $0.1 \, \mathrm{Mpc} < r_c < 1 \, \mathrm{Mpc}$.

The final redshift probability distribution is obtained by assuming:

 \begin{itemize}
     \item The variation of the galaxy bias within the clustering range of the reference set is negligible so that $\bar{w}_{rr}(r_c,z)=b^2_r(z) \bar{w}_{DM}(z)  /\Delta z$.
     \item Within the redshift range $\Delta z$, the relative variation of $\frac{d P}{d z}(z)$ dominates over $\bar{b}_t(z)$ and we approach the regime where $\frac{d P}{d z}(z) \rightarrow P(z)\delta(z-z_0)$.
 \end{itemize}
 
\begin{align}
P_{m,z} \propto \frac{\bar{w}_{tr}}{ \sqrt{\bar{w}_{rr} \Delta z}}\times \frac{1}{\overline{b}_t(z)\sqrt{ \bar{w}_{DM}(z) }} ~ .
\label{eq:Pz}
\end{align}

In words: Eq. \ref{eq:Pz} shows that the redshift distribution for a target sample within a magnitude bin at a specific redshift $P_{m,z}$ depends on the cross-correlation amplitude $\bar{w}_{tr}$ between the target and the reference sample, the auto-correlation amplitude $\bar{w}_{rr}$, the bias of the target sample $\bar{b}_t(z)$ and the dark matter clustering amplitude $\bar{w}_{DM}$.

For the measurement of the cross-correlation amplitude $\bar{w}_{tr}$ and auto-correlation amplitude $\bar{w}_{rr}$ we used the estimators by \cite{Peebles1974} and \cite{Landy1993}, respectively. The pair-counts needed in both estimators were measured using the python package \texttt{corrfunc} \citep{corrfunc2017}.

Using Eq. \ref{eq:Pz} we can calculate the redshift probability distribution of the target sample up to an unknown normalisation which depends in detail on the unknown, and evolving, bias of the target sample (see Appendix for further tests). 

By calculating the cluster-$z$s according to Equation \ref{eq:Pz} for different magnitude-binned sub-samples of the target dataset, the necessary information for the determination of the GLF is obtained. We use a redshift binning of $\Delta z = 0.02$ and a binning in magnitude $\Delta \mathrm{mag}_{W1} = 0.25$. The redshift distribution of our reference set with these bins is shown in Fig. \ref{fig:redshift_distri}. To get an estimate of the uncertainty in the measurements, the errors are calculated via jack-knifing of 12 sub-sets with similar area and number counts.

Neglecting the variation of the galaxy bias of the target sample with redshift, the resulting cluster-$z$s are only proportional to the true redshift distribution. They are related to the true number distribution by the normalisation factor $A_m$, such that $P_{m,z} = A_m \times N_{m,z}$. This magnitude-dependent normalisation factor can be derived analytically by comparing the $P_{m,z}$ with a parametric GLF model as described below.

\subsection{Model parametrisation}
\label{sec:model}
As described above, the calculation of the cluster-$z$ results in a probability distribution $P(z|m)$, which is proportional to the GLF, $\phi(z|M) dV \propto P(z|M) /dz$, where $M = m - DM + 2.5\log(1+z)$.

To determine the normalisation factor $A_m$, which connects the probability distribution with the number distribution ($A_m = P(m,z)/N(m,z)$) we have to assume a model of the GLF. For this model we use a single Schechter function \cite{Schechter1976} as our model parametrisation,

\begin{align}
S(M|M^*,\alpha,\phi^*) =  & 0.4 \ln 10 \phi^{*}\left[10^{0.4\left(M^{*}-M\right)}\right]^{\alpha+1} \nonumber\\
&\times \exp \left[-10^{0.4\left(M^{*}-M\right)}\right] \mathrm{d} M
\end{align}

We add a simple parametrisation for the redshift evolution \citep{Lin1999}, using the parameters $Q$ and $P$ describing a linear evolution of the logarithmic galaxy density $\log(\phi_i^*)$ and characteristic magnitude $M^\dagger$. In this model, the slope $\alpha$ is kept constant with redshift,

\begin{align}
    M^\dagger(z)&=M^\dagger-Q \times z \\
    \phi^*(z)&=\phi^*_0 \times 10^{0.4 \times P \times z}\\
    \alpha(z)&=\alpha
\end{align}

Despite it being common in the literature to use a double Schechter function in the optical when fitting the GLF \citep[e.g.][]{Baldry2008, Moffett2016} we have decided to use a single Schechter parametrisation, as we are limited by the confusion limit of WISE and our data are not deep enough to justify the addition of a second Schechter component.

The absolute luminosity derived here is in the observers' frame and no k-corrections have been applied so far. Therefore, the parameters, especially $M^\dagger$, are within the observers' frame. Instead of correcting $M^\dagger$, we correct the resulting observers' frame GLF, which is described in Sec. \ref{sec:results}, giving us the opportunity to compare our GLF to other surveys in the absence of potential uncertainties due to k-corrections.

\subsection{Parameter estimation}
\label{sec:mcmc}
Given a set of parameters, this evolution model in combination with the single Schechter parametrisation provides a prediction for the $N(m,z)$. As the output of the cluster-$z$s is only proportional to the redshift distribution $N(m,z)$ a normalisation factor is needed. We obtain the normalisation factor 

\begin{align}
    A_m = \frac{\sum_z \Phi(z|m) \times P(z|m) / \sigma_{P(z|m)}^2 }{\sum_z \Phi(z|m) \times \Phi(z|m) / \sigma_{P(z|m)}^2 }
    \label{eq:A-fac}
\end{align}

by least squares, such that $A_m = P(m,z)/N(m,z)$. The normalisation factor $A_m$ is therefore the maximum-likelihood solution given the GLF model $\Phi$, based on a certain parameter combination of $M^\dagger,\phi^*_0,\alpha,Q\, \mathrm{and}\, P$, and the cluster-$z$s data $P(z|m)$ in a respective magnitude bin.

For the fitting process the likelihoods of the SDSS Main, SDSS LRG, BOSS and GAMA cluster-$z$s are calculated separately using their own $A_m$s and are combined to one cluster-$z$s likelihood
$\ln(\Like_{cluster-zs}) = \ln(\Like_{SDSS_{Main}}) + \ln(\Like_{SDSS_{LRG}}) + \ln(\Like_{BOSS}) + \ln(\Like_{GAMA})$. 

As the use of the normalisation factors $A_m$ leads to a degeneracy between the global normalisation of the model with the $A_m$s, we use the GAMA spec-$z$ sample to constrain $\phi^*_0$, so that the final likelihood becomes $\ln(\Like) = \ln(\Like_{cluster-zs}) + \ln(\Like_{spec-zs})$.  For the likelihood of the spec-$z$ sample $\Like_{spec-zs}$, we use a point-based likelihood function as described by \cite{Marshall1983}. In this approach sample variance is neglected and therefore our errors do not represent field-to-field variations.

For the total likelihood the cluster-$z$ likelihood as well as the spec-$z$s likelihood are combined and the parameters $M^\dagger, \phi^*_0, \alpha, Q \, \mathrm{and} \, P$ are maximized using the MCMC sampler \texttt{emcee} \citep{emcee2019}. Here uniform priors are used to sample the parameter space. In order to achieve convergence the fit is continued until the estimated auto-correlation time \citep{Goodman2010} is less than $\tau = N_{samples}/50$.
    
\section{Results}
\label{sec:results}
The calculation of the cluster-$z$s results in a probability distribution $P(m,z)$, which gets normalised into a number distribution $N(m,z)$ by the use of the normalisation factors $A_m$, as described above. By applying the appropriate distance scaling to the magnitude/luminosities, and normalising counts by volume, the GLF $P(m,z) dM$ follows directly from the counts $N(m,z)$.

As described in Sec. \ref{sec:kml} we translate the observers' frame GLF to rest-frame by applying k-corrections to the cluster-$z$s GLF results. With the second step of using mass-to-light ratios we obtain the GSMF. As the cluster-$z$s require k-corrections and mass-to-light ratios for ensembles of galaxies, we use two approaches: an empirical model as well as predictions from simulations to obtain the factors needed and compare the results based on these two approaches.

\subsection{Best fitting model and the recovery of the number distribution}
\label{sec:Nz}
\begin{figure}
    \centering
    \includegraphics[width=\columnwidth]{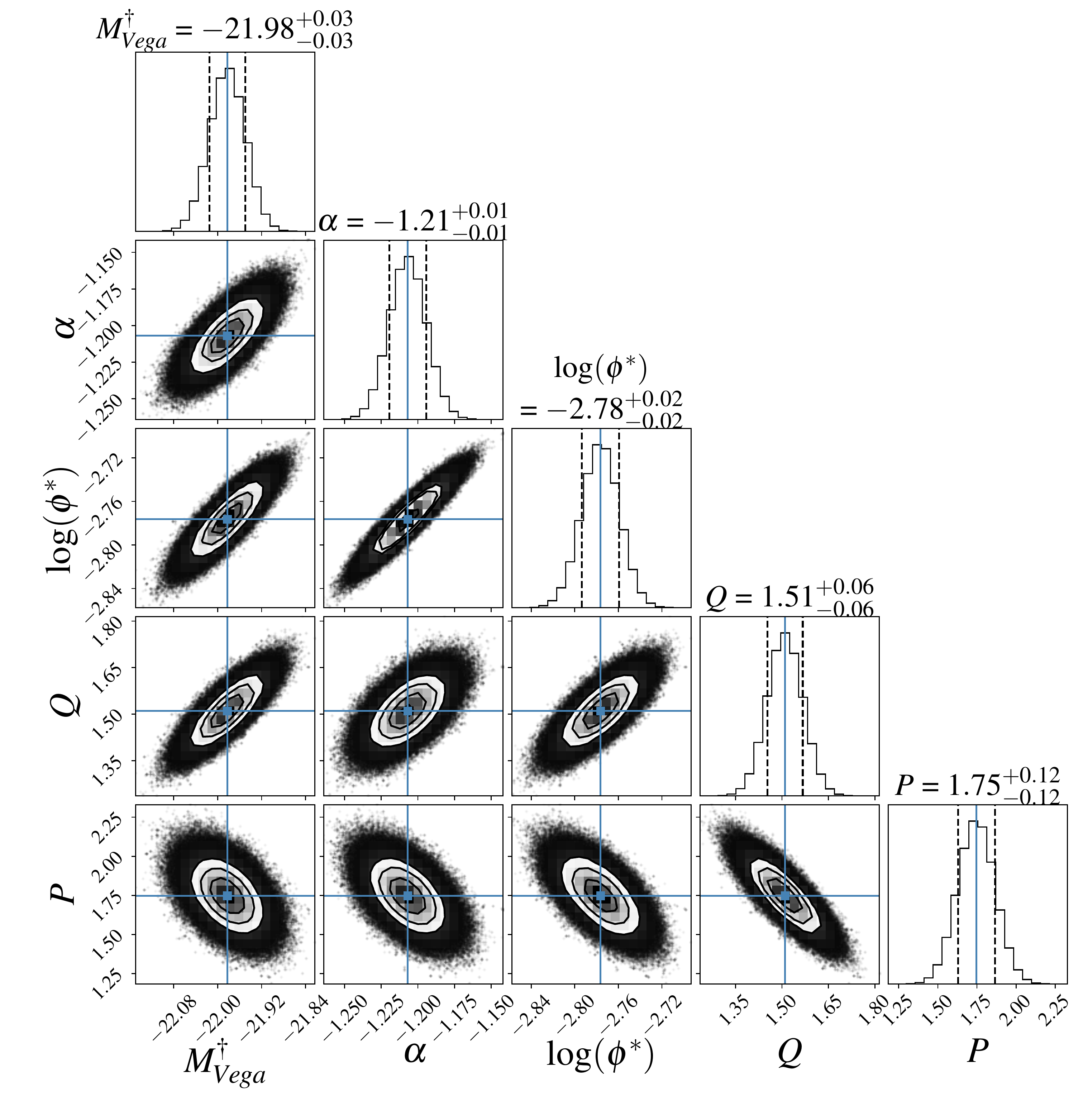}
    \caption{Resulting posterior distribution of the single Schechter function fit to the combined observers' frame cluster-$z$s and spec-$z$s likelihood described in Sec. \ref{sec:mcmc}.}
    \label{fig:PosteriorSingleSchechter}
\end{figure}

After optimizing the combined likelihood of the spec-$z$s and cluster-$z$s, we obtain the best fitting model, which we use to normalise our cluster-zs measurements as described in Sec. \ref{sec:mcmc}. The resulting posterior distribution is shown in Fig. \ref{fig:PosteriorSingleSchechter}, where the convergence of the fit can be seen from the gaussian posteriors on each parameter. As the fit is done in observers' frame values, the resulting parameters have to be treated as such. The GLF model shows a redshift evolution of $M^\dagger$ to brighter absolute magnitudes ($Q = 0.83\pm0.02$), while the galaxy density slightly increases with redshift ($P = 1.72\pm0.05$). The best-fit parameters including their uncertainties derived from the posterior distribution are displayed in Tab. \ref{tab:fitparam}.

\begin{table}
\centering
\resizebox{\columnwidth}{!}{\begin{tabular}{ccccc} \hline
$M^\dagger$ & $\alpha$ & $\log(\phi^*)$ & $Q$ & $P$ \\ \hline
$-21.98\pm0.03$ & $-1.21\pm0.01$& $-2.78\pm0.02$ & $1.51\pm0.06$& $1.75\pm0.12$\\
\end{tabular}}
\caption{Best-fit parameters including uncertainties of the single Schechter fit from the posterior distribution in Fig. \ref{fig:PosteriorSingleSchechter}.}
\label{tab:fitparam}
\end{table}

\begin{figure*}
    \centering
    \includegraphics[width=\textwidth]{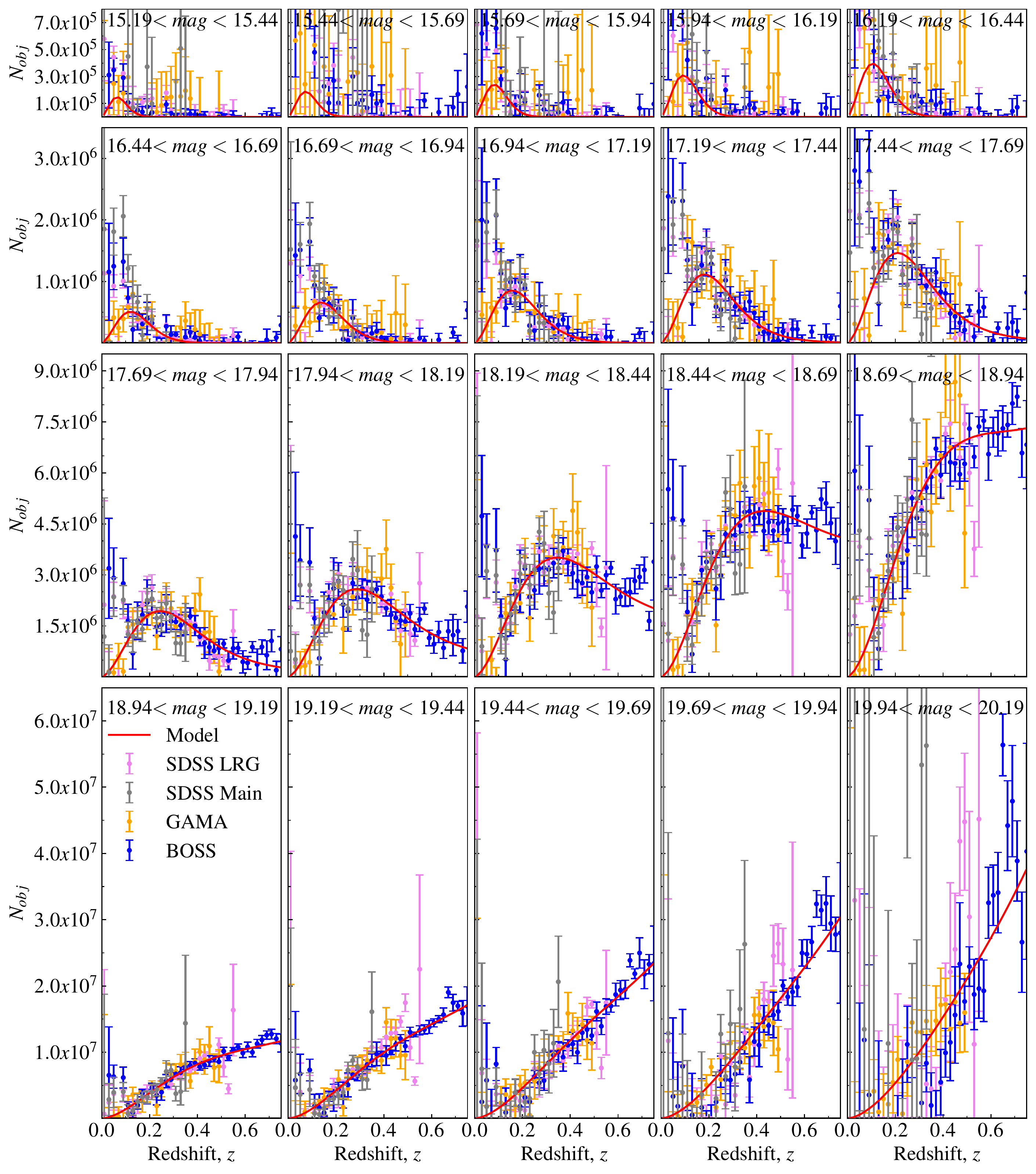}
    \caption{The resulting redshift distributions of our analysis. The BOSS data points are shown in blue, SDSS LRG in violet, SDSS Main in grey and GAMA in orange. The model is shown by the red line. Here all cluster-$z$s sets and the model are normalised such that they would cover $7350 \mathrm{deg}^2$ to be comparable. The $N(z|m)$ can be seen to increase with redshift, and while covering the whole redshift distribution of the target dataset for bright magnitudes, a significant fraction of the target data at fainter magnitudes is beyond the redshift range of our study.}
    \label{fig:Nz}
\end{figure*}

Using this best fitting model we normalise the different cluster-$z$s measurements using the A-factor described in Sec. \ref{sec:mcmc}. In Fig. \ref{fig:Nz} we compare the resulting $N(z|m)$ of each cluster-$z$s with the best fitting model. Over all magnitude bins it can be seen that the model is in good agreement with the cluster-$z$s. It can be seen how the galaxy distribution for the brighter bins is covered completely by our analysis while the end of the redshift distribution is beyond the redshift range of our study. By comparing the total number of galaxies expected from our analysis with the number of objects in the target catalogue with $\mathrm{mag}_{W1,\mathrm{Vega}} = 17.5$, we find that $\sim36\%$ of objects are within $z<0.6$ and $\sim53\%$ within $z<0.75$. This is in close agreement with the $52\%$ at $z<0.75$ found in \cite{Jarrett2017} using the same magnitude limits.

The target density of each reference set is the main limitation to the cluster-$z$s. While BOSS can produce reasonable results to $z\lesssim0.7$ for most magnitude bins, SDSS Main is limited to $z\lesssim0.4$, SDSS LRG to $z\lesssim 0.55$ and GAMA to $z\lesssim 0.5$. As the cluster-$z$s from BOSS are becoming increasingly uncertain at $z\gtrsim 0.6$ we decided to limit our measurement of the GLF to $z<0.6$, similar to \cite{Dai2009}.

As mentioned before, the resulting cluster-$z$s are broadly in agreement but show differences due to noise. While the cluster-$z$s from all sets show agreement with the model at intermediate redshift, especially at low redshift the cluster-$z$s are overestimating the model redshift distribution. This is especially true for the SDSS Main cluster-$z$s as its shape is incompatible with a Schechter-like $\Phi(M)$. While the uncertainties of the measurements might be underestimated (e.g. we neglect sample variance) the overestimation could be a result of the unknown galaxy bias, or issues with the WISE point-source photometry at low redshift. As galaxies at low redshift are well-resolved in the W1-band \citep[][]{Jarrett2013, Jarrett2017, Cluver2020}, the point-source photometry is missing flux, resulting in a difference of up to $\sim0.5$ mag, which makes the unWISE point-source photometry at low redshifts untrustworthy (see Appendix \ref{app:photometry} for more details). The uncertainties of these low redshift data points are an additional indicator that these data points are driven by noise and potential statistical fluctuations in the cluster-$z$s.

\begin{figure}
    \centering
    \includegraphics[width=\columnwidth]{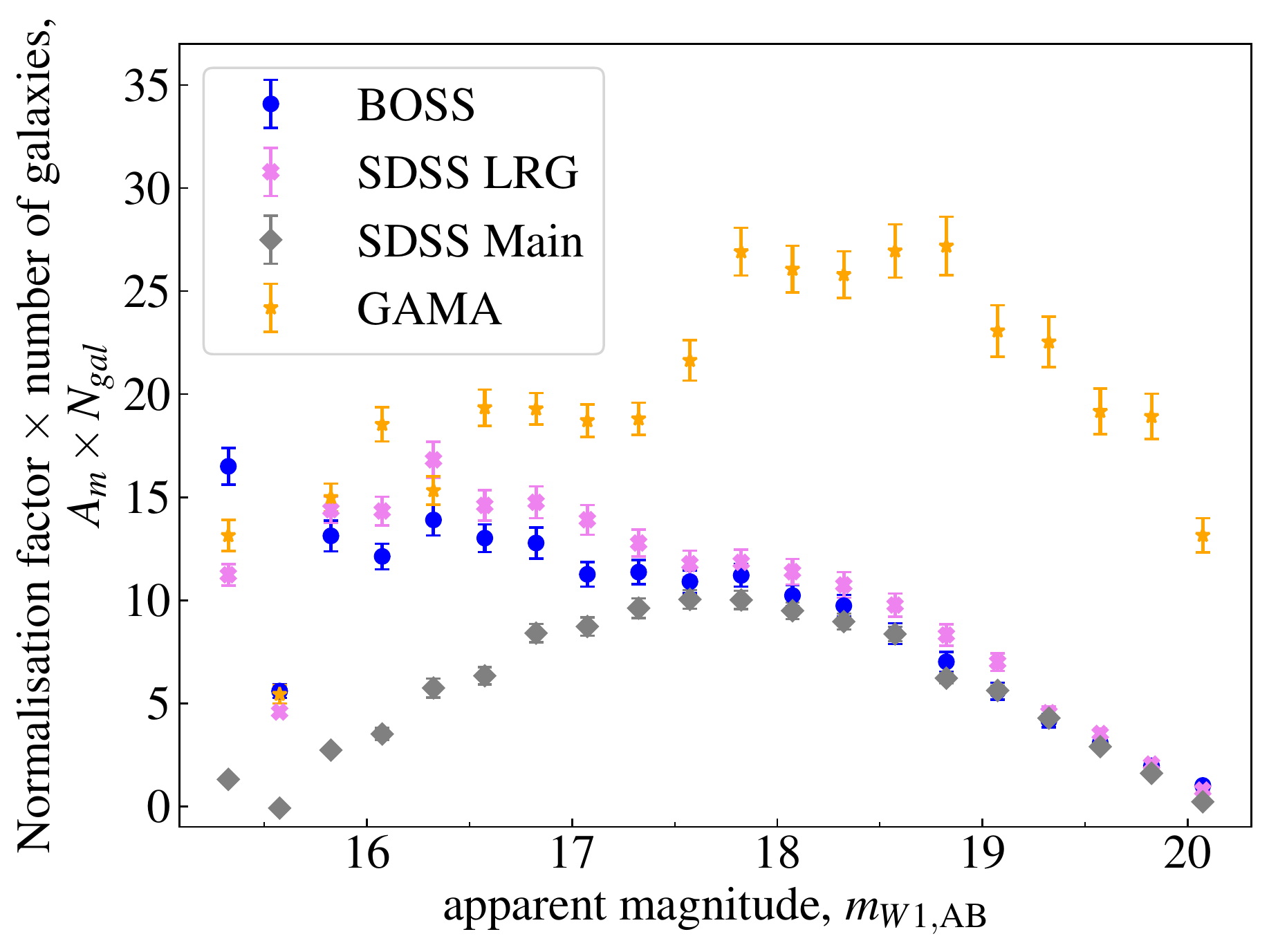}
    \caption{The normalisation factor $A_m$ for each reference set for each magnitude multiplied by the number of target objects. The $A_m$ from GAMA show larger values than the other surveys, which is a result of the much smaller area of GAMA compared to SDSS and BOSS. SDSS Main, SDSS LRG and BOSS are generally in good agreement at faint magnitudes, but for brighter bins the $A_m$s SDSS Main tend to result in lower values than the other sets.}
    \label{fig:Afac}
\end{figure}

By investigating the individual A-factors in Fig. \ref{fig:Afac}, it can be seen that the A-factors for SDSS Main follow a different trend than the A-factors of BOSS and SDSS LRG. While for faint magnitudes the different sets are in agreement, at bright magnitudes the scaled SDSS Main A-factors result in smaller values than the A-factors of BOSS and SDSS LRG. While the origin of this behaviour is not clearly understood, it provides further indication that the results from SDSS Main are noisy. 

The A-factors for GAMA are different to the other reference sets due to its different characteristics to the other dataset. The variation within the GAMA cluster-$z$s is larger than for the other samples at all magnitudes. As GAMA is covering a much smaller area than SDSS or BOSS, this shows the importance of using large areas for this study to reduce the impact of cosmic variance. To see how these differences between the sets influence the resulting GLF, we derive the GLF for each set individually in the next section.

\subsection{The galaxy luminosity function}
\begin{figure*}
    \centering
    \includegraphics[width=\textwidth]{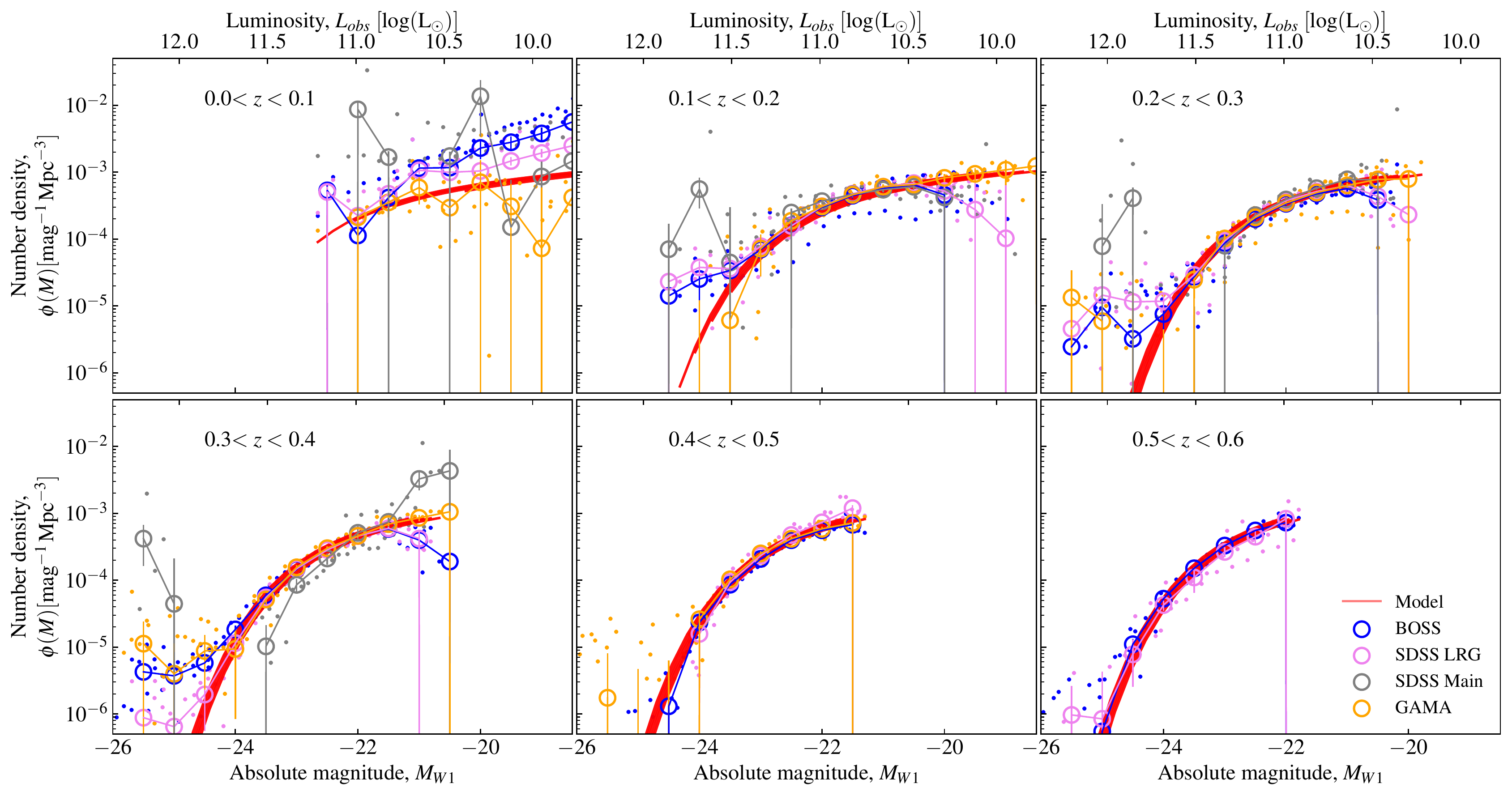}
    \caption{Comparison of the derived observers' frame GLF for the GAMA (orange), SDSS Main (grey), SDSS LRG (violet) and BOSS (blue) cluster-$z$s and the best fitting model (red) in six redshift bins with $\Delta z=0.1$. It can be seen that the difference in the GLF between the different cluster-$z$s is small apart from the lowest redshift bin. In this figure all measurements are shown, while we exclude data from BOSS at $z<0.2$ and from SDSS Main at $z<0.1$ for the further analysis. As a reminder to the reader: this diagram displays the observers' frame GLF and not the rest-frame. For any rest-frame comparisons the resulting GLF has to be shifted with wavelength given its redshift. E.g. the W1 flux observed in the last bin at $z>0.5$ is close to the rest-frame K-band and should consequently not be compared to rest-frame W1 results.}
    \label{fig:LFPz}
\end{figure*}

After normalising the cluster-$z$s using the A-factors, the GLF is derived by normalising the $N(z|m)$ (Fig. \ref{fig:Nz}) by the corresponding cosmological volumes. The resulting GLF without k-corrections, but including band-pass stretching, is shown in Fig. \ref{fig:LFPz} for each of the different cluster-$z$s. Here the individual data points for each cluster-$z$s are shown and the mean values within $0.5$ mag bins are over-plotted using large circles and error bars. The red line shows the model using the best fitting parameters from Tab. \ref{tab:fitparam}. At $z<0.1$ the results tend to result in larger values than the model. While the results from GAMA, BOSS and SDSS LRG provide clear trends, the results from SDSS Main are dominated by noise. At higher redshifts the resulting GLFs from all the different cluster-$z$s agree with each other and with the model.

At the bright end of the GLF the cluster-$z$s are able to trace the exponential cut-off down to a sensitivity of $\sim 10^{-5} Mpc^{-3}$ at low redshift. At higher redshift the sensitivity increases to $\sim 10^{-6} Mpc^{-3}$. As expected from Fig. \ref{fig:Nz}, the GLFs from the cluster-$z$s are overestimating the model at $z<0.1$ for all datasets (see discussion in previous section). This is especially true for the BOSS and SDSS data, while the GAMA is in better agreement with the model GLF. As the GLF and $N(z)$ from the SDSS cluster-$z$s are suffering from large scatters we exclude the data points from SDSS Main at $z<0.1$ for further results. As BOSS is affected due to its bright limit of $\mathrm{mag}_r > 16$ \citep[][]{Alam2017}, we also restrict our final results for BOSS to $z>0.2$.

\begin{figure*}
    \centering
    \includegraphics[width=\textwidth]{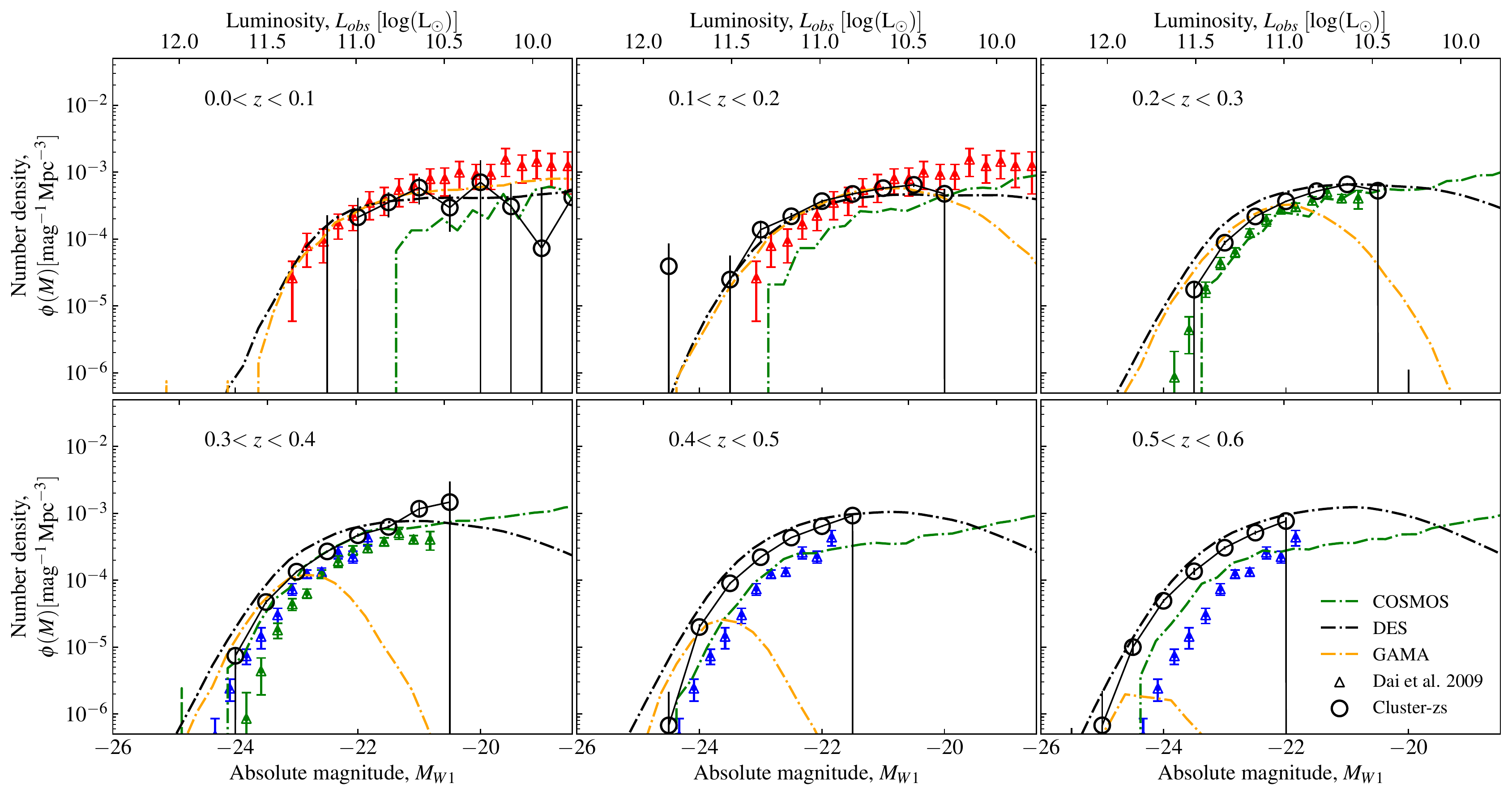}
    \caption{The measured GLF in observers' frame (black circles) in six $\Delta z=0.1$ redshift bins compared to data from the GAMA (orange line), DES (black line) and COSMOS (green line) surveys. The GAMA GLF is based on spec-$z$s, while DES and COSMOS are based on photo-$z$s. In addition the results from \protect\cite{Dai2009} are shown as triangles with errorbars. The results from \protect\cite{Dai2009} are colored according to their redshift: $z<0.2$ in red, $0.2<z<0.35$ in green and $z>0.35$ in blue.}
    \label{fig:LFnonkcorrComp}
\end{figure*}

Applying these redshift cuts to the cluster-$z$s we calculate the final observers' frame GLF by combining all the remaining data points, weighted by their respective survey area, and calculating the weighted mean. The resulting GLF is compared to similar measurements using the Dark Energy Survey (DES) \citep[][]{Gatti2018}, the Cosmic Evolution Survey (COSMOS) \citep[][]{Laigle2016}, GAMA and \cite{Dai2009} in Fig. \ref{fig:LFnonkcorrComp}. We correct the IRAC1 magnitudes from COSMOS and \cite{Dai2009} by calculating the difference in magnitude between IRAC1 and W1 using the galaxy SED templates from \cite{Brown2014} which results in correction factors between $-0.016$ and $-0.14$ magnitudes depending on redshift.

Our measurements in the lowest redshift bin are in agreement with the GLF from DES , GAMA and \cite{Dai2009} up to $M_{W1} \sim -20$, where the uncertainties become significantly larger. COSMOS at this redshift is, due to its very small surface area of $2 \mathrm{deg}^2$, suffering to large sample variance. At higher redshifts, the sample variance error decreases owing to the increasing volume, and COSMOS shows its full potential by probing deeper than all measurements. The best agreement of our GLF is with the GAMA GLF, while the GLFs from DES and COSMOS are over- and underestimating the GAMA GLF in most redshift bins, respectively. It can be seen that the GLF measurements based on IRAC1 magnitudes (COSMOS and \cite{Dai2009}) underestimate the GAMA, DES and our measurements and we assume that the correction from IRAC1 to W1 based on the SED templates is not capturing the complete offset. As GAMA is based on spec-$z$s, while DES is using phot-$z$s, this difference in the redshift derivation is potentially the reason for the differences in the GLFs of the different surveys. 

\subsection{Relating observed and intrinsic quantities}
\label{sec:kml}
We have measured the GLF as a function of the absolute luminosity as measured in the observers' frame band-pass, i.e. scaled for distance but not k-correction. This means that some of the evolution observed within the GLF is due to the W1 filter sampling different wavelengths at different redshifts.

The principal strength, and weakness, of our analysis, is that we never consider the redshift of any one particular source. Similarly, we only consider one wavelength and doesn't use any SED information. This means that we have to correct our observers' frame GLF by a population average if we want to obtain intrinsic properties. The difficulties within this approach are random scatter with the stellar population at fixed magnitude, and systematic variations in stellar populations as a function of magnitude and redshift.

This leaves two possible avenues: the first is to apply a simple empirical prescription of these effects. The advantage of using this approach is that it is transparent and simple to apply. The disadvantage is that empirical relations don't cover all the evolution with redshift and magnitude. The second avenue is to rely on a model of galaxy evolution from simulations. The advantage of this approach is that it is possible to capture the full complexity of galaxy evolution including random errors. The disadvantage is that the resulting properties will be affected by systematics within the simulations and the uncertainty about how well the simulation is able to mimic the true galaxy evolution in all its details.

\subsubsection{Rest frame corrections}
\label{sec:kcorr}
To derive the rest-frame GLF, we have to apply k-corrections $K$ to the observers' frame GLF. Here $K$ relates the apparent magnitude $m$ to the absolute magnitude $M$ of an object via the distance modulus $DM$, the extinction due to band-pass stretching $2.5\log(1+z)$, so $m = M + DM - 2.5\log(1+z) - K$ \citep{Wilson2002}.

Due to variations in stellar populations as a function of redshift and mass, the derivation of the k-corrections is a complex task which is different for each object. As we are not looking at each object individually we have to find k-corrections, which are representative of the whole population.

The first corrections we use are from an empirical model by \cite{Jarrett2023}, which is calibrated to real data but treats all galaxies according to their average behaviour. These corrections are based on the "Sb" galaxy template from \cite{Brown2014} as it provides a good mean for all galaxy types. In \cite{Jarrett2023}, the k-correction is described as an exponential function depending on redshift $z$ with $K = 1 - a_0 e^{\alpha z}$, where $a_0=1$ and $\alpha=-2.614$.

For the second approach, we use simulated k-corrections based on galaxy SEDs generated using \Mentari \, \citep[][]{Triani2023}. The \Mentari \, tool uses simulated galaxies calculated from the Millennium simulation \citep{Springel2005} and the semi-analytic model \texttt{Dusty SAGE} \citep[][]{Triani2020b}. Here \Mentari \, focuses on simulated galaxies with stellar mass $M^* > 10^8 M_\odot$ and generates SEDs from far-ultraviolet to far-infrared wavelength which includes stellar emission, dust attenuation and re-emission using the population synthesis code by \cite{Bruzual2003}. The details about the derivation of the k-corrections and their comparisons can be found in the appendix.

\subsubsection{Mass-to-light ratios and the galaxy stellar mass function}
\label{sec:MLratio}
\begin{figure*}
    \centering
    \includegraphics[width=\textwidth]{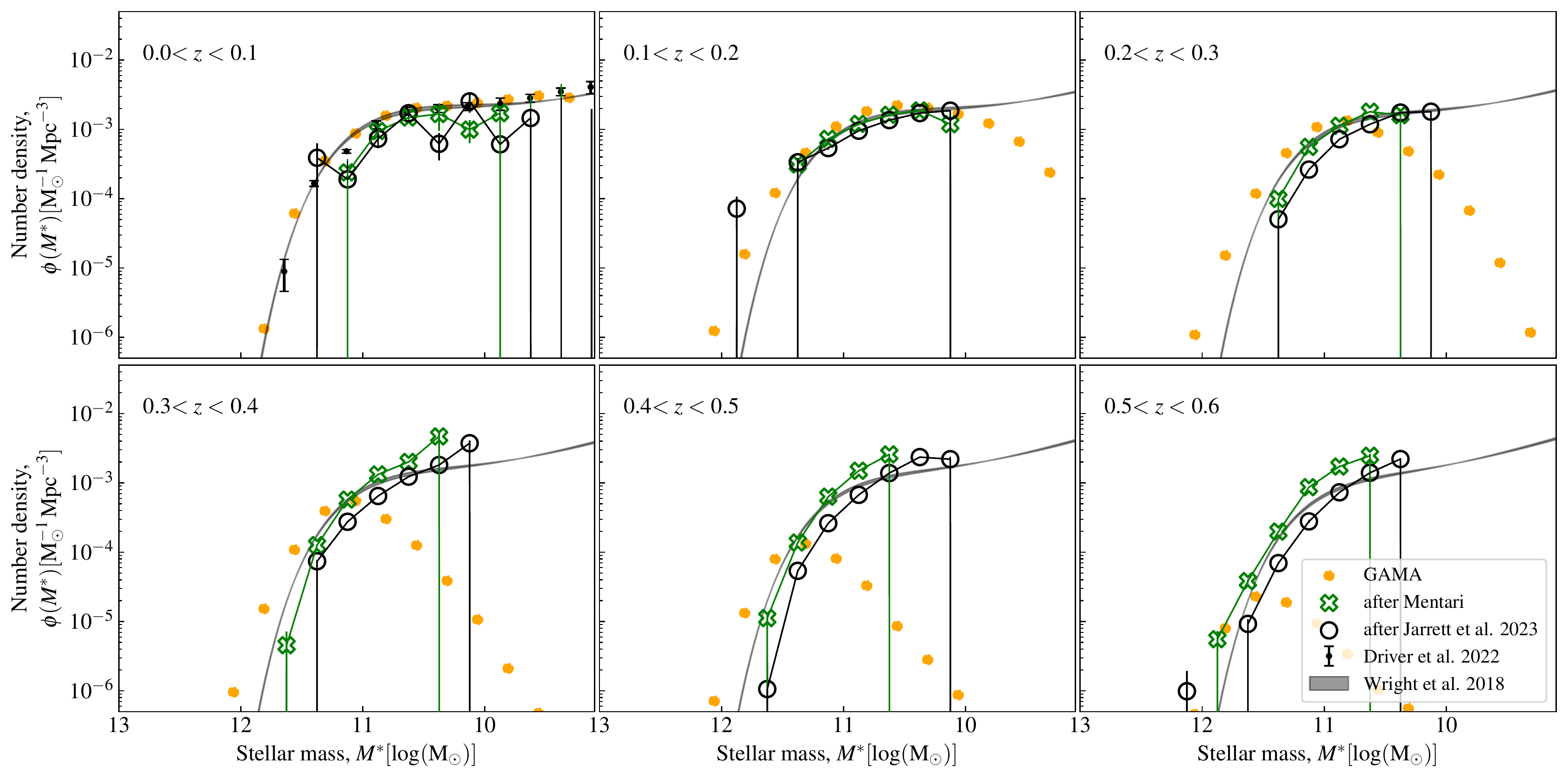}
    \caption{ The GSMF based on k-corrections and mass-to-light ratios by \protect\cite{Jarrett2023} (black) and \Mentari \, (green). In comparison the results by \protect\cite{Wright2017} are shown in grey as well as the results from \protect\cite{Driver2022} at $z<0.1$ as black scatter points. In addition the number counts from the GAMA survey are shown as orange points. Here the differences between the two methods can be seen clearly (see Sec. \ref{sec:DisskcorrMLratio} for further discussion).}
    \label{fig:MF}
\end{figure*}

For the derivation of mass-to-light ratios, we use a similar approach as for the k-corrections. Again we use an empirical relation from \cite{Jarrett2023} of the form $\log (M^*) = 1.2077 + 0.6092 \times \log (L_{W1}) + 0.0220 \times \log (L_{W1})^2$, where $L_{W1} = 10^{0.4(M_{bol,\odot, W1}-M)}$. The empirical relation from \cite{Jarrett2023} is the result of a least-squares fit using W1 GAMA data. As our second approach, we derive the mass-to-light ratios from \Mentari \, which we apply in a similar manner. Both mass-to-light ratios are described in more detail in the appendix.

The two resulting GSMF based on the empirical relations from \cite{Jarrett2023} and \texttt{Dusty SAGE} in combination with \Mentari \, are shown in Fig. \ref{fig:MF}. For the derivation of the GSMF the k-corrected GLFs of each approach are combined with their corresponding mass-to-light ratios. While the difference between the two GSMFs is rather small at low redshift, the differences between the k-corrections and mass-to-light ratios lead to larger offsets at higher redshift. In general the GSMF from \Mentari \,indicates higher densities of galaxies compared to the GSMF using the model by \cite{Jarrett2023} and the comparison of the two approaches shows the range of potential results depending on which models are used.

\section{Discussion}
\label{sec:discussion}
We have laid out a process for using cluster-$z$s as a basis for measuring the near-infrared GLF and GSMF. and it has shown good qualitative agreement with earlier measurements using photo-$z$s and spec-$z$s. While statistical uncertainties are small, we are quite likely dominated by systematic effects.

In the following we will discuss potential sources of systematic error in our analysis: addressing the ranges and limits chosen during the calculation of the cluster-$z$s, the dependency of our experiment on the model parametrisation, the impact of the unknown bias evolution of the target dataset, potential bias and incompleteness of the anchor dataset, the uncertainties at low redshift and the differences of the k-corrections and mass-to-light ratios used. Thus we highlight key aspects for consideration in future studies.

\subsection{Cluster-z binning, ranges, and limits}
For the calculation of the cluster-$z$s three main decisions have to be made: the magnitude binning of the target set, the redshift binning of the reference set and the clustering ranges. The magnitude binning of the target dataset has to be set large enough such that each bin contains enough data points to have enough statistical power to obtain a meaningful redshift distribution. Too narrow bins would result in large uncertainties and scatter for the redshift distribution over all redshifts. This effect is most likely to occur in the brightest bins, due to low number counts. We have therefore chosen a magnitude binning where the number of target data points is significantly larger than the number of reference points for each bin.

The choice of $\Delta z$ is a trade-off between resolution and uncertainties. While too wide redshift bins result in limited information about the $N(z)$, too narrow bins result in large uncertainties and scatter in redshift bins with small numbers of reference points. Through testing of multiple redshift binning between $0.01 < \Delta z < 0.05$ we found that $\Delta z = 0.02$ is a good trade-off for the redshift resolution for our experiment.

For convenience and for better comparison with other studies we have chosen to use regular grids for our binning. In future studies flexible bins with similar statistical power could be used instead, as this would allow to obtain more information especially at the faint end.

While the binning defines the final resolution of our experiment, the choice of clustering-ranges $r_c$ is much more important, as it defines the amount of information retrieved from the cross-correlation. Here the lower limit has to be large enough to exclude potential self-correlations and avoid fibre collision, while not missing genuine associations. As it has been shown by \cite{Gordon2018}, the angular correlation functions deviate for a sample limited by fibre collision compared to an unbiased sample at separations below the fibre collision limit. Thus we have excluded correlations on smaller scales then the fibre collision limit from the cluster-$z$s calculation. While the lower limit is a hard limit, the upper limit for $r_c$ is more flexible. It has to be large enough to capture the LSS, but also not too large, such that the assumption of linear bias is till valid and statistical noise added by uncorrelated background galaxies is limited. We have performed multiple tests and found that our results are robust to small changes in $r_c$.

\subsection{Model parametrisation}
\label{sec:DisModelParam}
As discussed in Sec. \ref{sec:mcmc}, we need a parametric model to normalise the raw cluster-$z$s. While it is necessary to choose a model, the exact form of this model seems to be unimportant as in our experiment the model is only a tool to derive the scalar factors necessary to normalise the cluster-$z$s ($P(z|m)s$). The model therefore only provides the mean for a self-consistent description of the $P(z|m)$s, which is also the reason we don't focus on the best-fitting parameters of the model. During this study a double Schechter function was also considered and we found that the second component of the double Schechter is not well constrained. This shows the limited depth of the unWISE data and is the reason we have chosen to use a single Schechter fit.

\subsection{The unknown bias and its evolution}
The evolution of the target bias is the largest systematic in the derivation of the cluster-$z$s \citep[see][]{Menard2013}. In this study we assume a flat bias, which corresponds to an evolution of the target sample galaxy bias over redshift which balances the growth of the dark matter structure. Such a flat bias, as well as bias evolution in magnitude, would only affect the global amplitude of the $P(z|m)$s, which is included in the normalisation scalar of the A-factors. Variations in the target bias with redshift on the other hand would lead to a skew in the resulting redshift distribution with redshift, which is not addressed by the $A_m$s. A different bias evolution with redshift would therefore result in a change to the galaxy density with redshift. While this evolution might be partially absorbed into the density evolution parameter $P$, it still remains a potential source of uncertainty. 

To obtain an indication of the strength of the unknown bias we have measured the target bias over redshift using data from the Theoretical Astrophysical Observatory (TAO) \citep[][]{Bernyk2016} as well as with GAMA WISECat \citep[][]{Cluver2014} (see Appendix C for more details). We model the bias using a simple parametrisation of $\bar{b}_u \propto (1+z)^B$ and we find that the bias evolution of TAO and GAMA from $z\sim0$ to $z=0.6$ is to a first order independent of magnitude. The values for the redshift evolution of the bias for TAO and GAMA is represented by factors of $B \sim 1.5$ and $B \sim 2$ respectively, which results in a change of $\Phi$ of about $\sim 2 - 2.5$. This would result in a decrease in the density evolution $P$ of $\sim 1.25 - 1.65$ and therefore a change of $\Delta \log(\phi^*) \lesssim 0.4 $ at $z=0.6$.

\subsection{The spectroscopic anchor and the WISE point-source photometry}
As the cluster-$z$s themselves are not able to constrain $\phi^*_0$, we use spec-$z$ data from the GAMA survey as our $z\sim0$ anchor. Using this anchor can potentially lead to systematic errors caused by inconsistencies and mismatches. Especially errors in $\phi^*_0$ due to an incomplete or biased anchor would have an direct impact on the final GLF and GSMF. We have therefore applied strict magnitude and redshift cuts to the dataset to achieve completeness within our selection. Another systematic which could influence $\phi^*_0$ is that the small GAMA regions are prone to field-to-field variations. This variance has been measured by \cite{Driver2022} and we correct our results for this effect to a first order.

The point-source photometry from WISE is systematically missing flux, especially for brighter and low redshift galaxies. By comparing the profile vs aperture fitted photometry of WISECat from GAMA in the Appendix, we find that a change in measuring the flux results in an $\gtrsim 0.2$ mag offset between the two methods. The effect of the missed flux is balanced by the $A_m$s to a first order, as the normalisation doesn't depend on the number of target objects per magnitude bin. The differential effects, which will "tilt" the GLF and GSMF at fixed redshift or impact the inferred redshift distribution at fixed magnitude, are of larger concern. As shown in the Appendix the mean fraction between the profile and aperture redshift distributions is $\frac{N_{profile}(z)}{N_{aperture}(z)} \sim 0.9$ with a peak-to-trough ratio of up to a factor of $2$. By fitting a power-law to the $N(m|z)$ ratios at the lowest redshift bin we find that a correction for the missed flux would steepen the slope by $\Delta \alpha \sim -0.1$. The change in $\alpha$ has to be seen as an upper limit as the difference between the two photometries becomes smaller with redshift. These findings show the limitations of our study at low redshift and highlight the importance of good photometry. 

\subsection{Uncertainties at low redshift}
We have seen that there are apparent inconsistencies in the cluster-$z$ results at $z \lesssim 0.1$: in particular, the cluster-$z$ results derived using the SDSS Main and to a lesser extent the SDSS LRG samples are significantly and systematically high in the first redshift bins. While we have not been able to determine the nature or cause of this issue, we describe some of the potential error sources and tests we have performed to minimize their impact in the following.

As the overestimation is occurring at low redshifts only, the unknown bias cannot be the source for this behaviour (see previous section for more detail about the unknown bias). It is also unlikely that confusion or blending is the reason for this behaviour, as it is occurring at the wrong scale. Another idea is that missed flux is the source of this problem, but additional tests showed that the missed flux is not resulting in these offsets (see Appendix A). The fact that different reference sets are in tension at low redshift shows that these results are not robust. Especially the difference between SDSS and the other surveys suggests, that the source of this overestimation lies within unaccounted systematics of the reference sets, and especially within SDSS. One potential source could be the selection function of the surveys used. We tried to address this issue by using the random datasets and weightings, where available, provided by the surveys.

Having the limitations of the SDSS cluster-$z$s at $z<0.1$ in mind, the results from different reference sets with quite different characteristics agree well at higher redshift. This suggest that these results at higher redshift are robust and only the lowest redshifts are dominated by noise. It shows that we are able to handle the different biases in the different reference sets and it is underlining the value of analysing different reference sets independently and checking for consistency in the results.

\subsection{K-corrections and mass-to-light ratios}
\label{sec:DisskcorrMLratio}
As we are not deriving redshifts for individual galaxies, we are only able to use k-corrections and mass-to-light ratios for a statistical ensemble. The challenge is therefore to account for multiple variations of stellar populations as function of mass and redshift. A model capturing this variety is therefore needed to derive these properties. This can be achieved by two broad approaches, either empirical or theoretical. The advantage of the empirical model is that it is using templates based on real measurements, which is also its limitation as the data used can be biased and incomplete. In the case of an theoretical approach the information of the galaxies within is as complete as the models allow, but it is unclear how well the model is able to mimic the true galaxy properties in every detail.

Astrophysically, one of the main uses of the GLF and GSMF is the comparison to models and simulations. While significant theoretical work is undertaken to improve the models, it might be beneficial for the observer to focus on the observers' frame GLF instead of intrinsic values, as the modeller must consider the stellar populations of each galaxy in their simulation already. While using the predictions from \Mentari \, to derive the k-corrected GLF and GSMF, we could also compare the observers' frame GLF from \Mentari \, to our measurement as this would provide the same information. As this would also allow to test for different models, it might be beneficial to define the observers' frame GLF and raw number counts as the meeting point between our observations and modellers. This is of increasing importance, especially when the GLF is known in multiple wavebands.

\section{Summary and Conclusion}
\label{sec:summary}
The aim of this study has been to infer the evolving near-infrared GLF and GSMF to the limits of the unWISE photometric catalogue, based on clustering redshift inference. Our motivation for this work has been two-fold. Firstly the analysis of WISE has been limited by the availability of spec-$z$s (e.g. GAMA) or photo-$z$s (e.g. DES), which creates joint IR and optical selection effects. Secondly we aimed to create a proof of concept as we are looking ahead to the new wide and extremely deep imaging and photometry surveys such as e.g., LSST \citep[][]{Ivezi2019}, Roman \citep[][]{Akeson2019}, and Euclid \citep[][]{Racca2016}.

The principal problem in our analysis is the relation between the unnormalised cluster-$z$s within each magnitude bin ($P(z|m)$) and the corresponding number distribution. After calculating $P(z|m)$ we use a simple parametric model of the evolving GLF to derive the normalisation factors $A_m$. Thus also an external constraint on $\phi^*_0$ is needed, for which we use spec-$z$s data from GAMA, which is described in Sec. \ref{sec:mcmc}.

The calculation of the cluster-$z$s is limited to the availability and properties of reference sets, as different reference sets probe different regimes in volume, mass and redshift. One novel aspect of our analysis is the combination of multiple reference sets to get the most information from the data. This enables us to probe different spatial scales for different samples.

As basic results we show the resulting observers' frame GLFs for each reference set in Fig. \ref{fig:LFPz}. We see good agreement between the GLFs for reach reference set at $z \gtrsim 0.1$. For the lowest redshifts, there are strong tensions between the results, especially with SDSS Main, which we are unable to explain. While this tension points to some unidentified issue in our low-z results, the good agreement for $z > 0.1$ says that these results are more robust.

By comparing the observers' frame GLF with results from GAMA, DES, COSMOS and \cite{Dai2009} in Fig. \ref{fig:LFnonkcorrComp} we have seen that our measurements are in general in good agreement with these surveys and studies.

In Sec. \ref{sec:discussion} we explore potential sources of systematic uncertainties. In particular, we highlight that the missed flux changes due to the point-source photometry induces a change in the slope by $\Delta \alpha \sim -0.1$ and a correction for the evolution of the unknown bias changes the density evolution by $\Delta P \sim -1.25 \, \rm to -1.65$. As the change in $\alpha$ increases the number density especially at the faint end, the change in $P$ reduces the number density with redshift. Applying both of these corrections results in almost no difference at low redshift and a slightly shallower observers' frame GLF at high redshift.

Another complication of our method compared to conventional photo-$z$ or spec-$z$ approaches are k-corrections and mass-to-light ratios. We have applied two approaches to get an estimate of the plausible range of the resulting GSMF and it shows that the application of k-corrections and mass-to-light ratios leads to an additional uncertainty due to the different results between the models. While this effect can in principle be calibrated, we suggest it is beneficial to consider the observers' frame GLF compared to predictions from models instead of rest-frame properties.

We have measured the GLF down to $\mathrm{mag}_{W1,\mathrm{Vega}} = 17.5 \, (30\mu \mathrm{Jy})$ within $z<0.6$. We have chosen this redshift limit, firstly as it is approaching the useful limit of the BOSS reference set. A proper treatment of these effects would require a more sophisticated GLF model. In this study we have accounted for $\sim36\%$ of unWISE galaxies within $\mathrm{mag}_{W1,\mathrm{Vega}} = 17.5$.

This approach can easily be extended to fainter magnitudes and to higher redshifts, with an appropriate reference set. This is most significant in connection with new imaging surveys like LSST, Euclid, or Roman in conjunction with spectroscopic surveys like 4MOST \citep[][]{Jong2019}, DESI BGS \citep[][]{Hahn2022} or WAVES \citep[][]{Driver2019}. Using this technique it will potentially be possible to probe down to $\sim10^{8} M_\odot$ up to $z\lesssim 1$ with LSST, assuming a completeness limit of $\mathrm{mag}_i \sim 27$, $\sim10^{10} M_\odot$ at $z\lesssim 2$ with Euclid at $\mathrm{mag}_Y, \mathrm{mag}_H \sim 24$ and $\sim10^{9} M_\odot$ at $z\lesssim 1$ with Roman at $\mathrm{mag}_{F184} \sim 27.4$. 

\section*{Data Availability}
The  derived  observers' frame GLF  measurements  shown  in  this  article  are available in the article and in its online supplementary material.

\section*{Acknowledgements}
We would like to thank the reviewer for his/her insightful comments and constructive remarks. GSK acknowledges financial support received through a Swinburne University Postgraduate Research Award. For this study \texttt{PYTHON} has been used for the data analysis, and we acknowledge the use of Matplotlib \citep[][]{Hunter2007} for the generation of figures in this paper.
\bibliographystyle{mnras}
\bibliography{sources} 

\appendix
\section{Missed flux in WISE point-source photometry}
\label{app:photometry}
\begin{figure}
    \centering
    \includegraphics[width=\columnwidth]{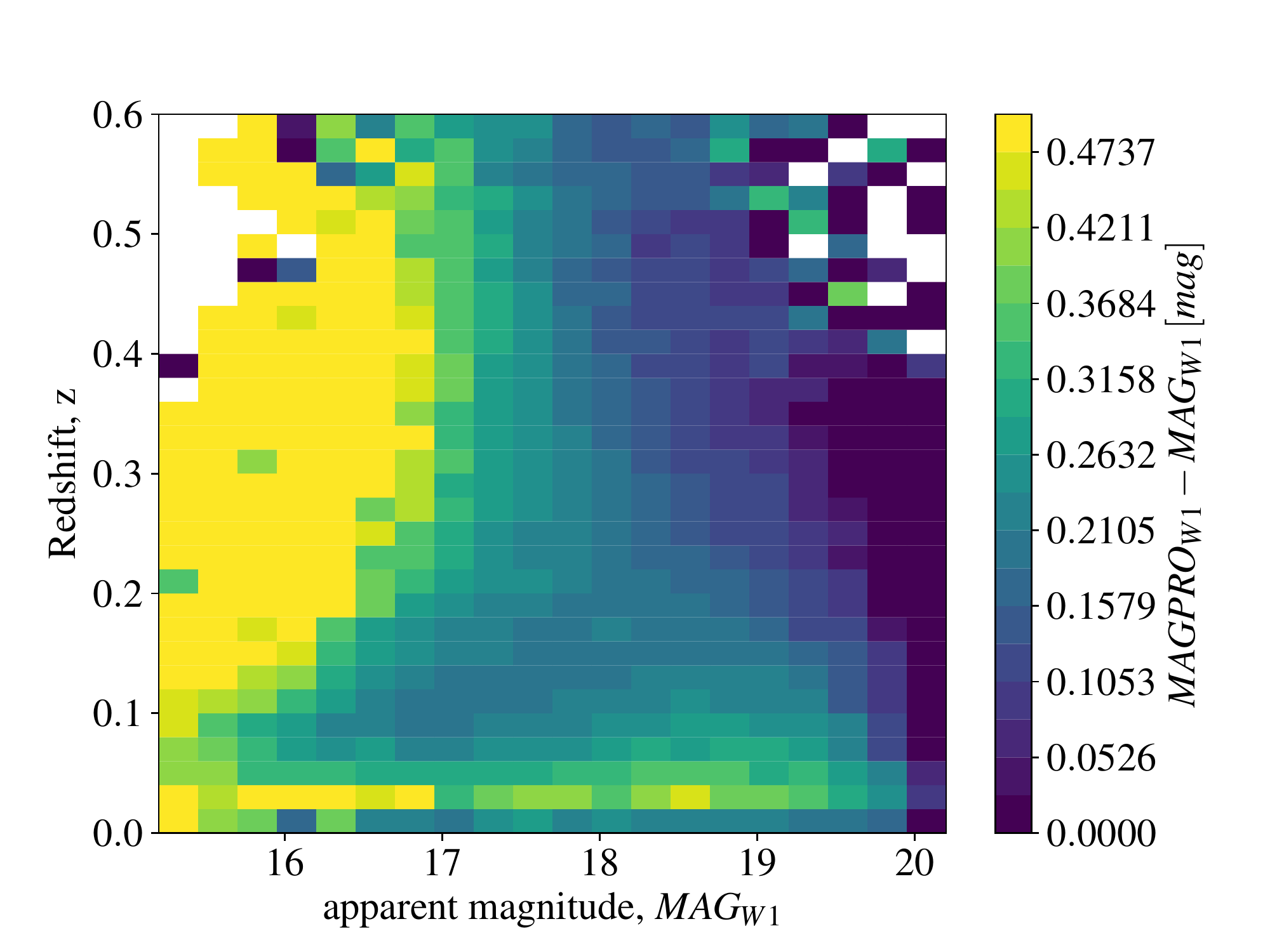}
    \caption{Difference between the resulting W1 magnitude measurements from GAMA using a profile fitting approach ($MAGPRO$) vs fitting apertures ($MAG$). The difference between the two different ways of measuring the photometry can be up to $\sim 0.5$ mag at low redshifts.}
    \label{fig:Photometry}
\end{figure}

Using point-source photometry leads to an underestimation of flux at low redshift and for bright and especially resolved galaxies. Although the W1-band doesn't have good spatial resolution, it is sensitive enough so that galaxies are not point sources and the flux from the "wings" can be captured using small apertures, which the PSF misses \citep[e.g.][]{Cluver2020}. To estimate the impact of the point source photometry we investigate the WISECat catalogue from GAMA \citep[][]{Cluver2020}. In WISECat the near-infrared photometry from ALLWISE \citep[][]{Cutri2014} has been measured for all objects in the GAMA equatorial regions detected by WISE. Here a profile fitting technique similar to WISE, as well as aperture fitting technique has been used, which allows us to compare the different measurements for each object. For aperture fitting in WISECat, isophotal apertures for extended resolved sources and "standard apertures" provided by ALLWISE for all other sources are used. Fig. \ref{fig:Photometry} shows the difference between the photometry from profile fitting ($MAGPRO$) and from aperture fitting ($MAG$). The difference between the two approaches shows a mean offset of about $\sim 0.2$ mag and a maximum of up to $\sim 0.5$ mag for the brightest galaxies at low-$z$. Fig. \ref{fig:Photometry} shows that the profile fitting approach used by WISE is missing significant amounts of flux, especially for bright galaxies at low redshift.

To test the impact of the missed flux we calculate the cluster-$z$s for the GAMA objects in the W1-band for two samples, one binned by $MAGPRO$ and one by $MAG$. Despite using the same dataset as our target and reference set, this is nevertheless not a circular test, as the data is split in magnitude for the target set and in redshift for the reference set. We obtain the $N(z,m)$s for the $MAGPRO$ and $MAG$ samples by A-factor normalising the cluster-$z$s as described in Sec. \ref{sec:mcmc}. Instead of using a modeled distribution to obtain the normalisation factors we use their true redshift distribution respectively for our test.

\begin{figure*}
    \centering
    \includegraphics[width=\textwidth]{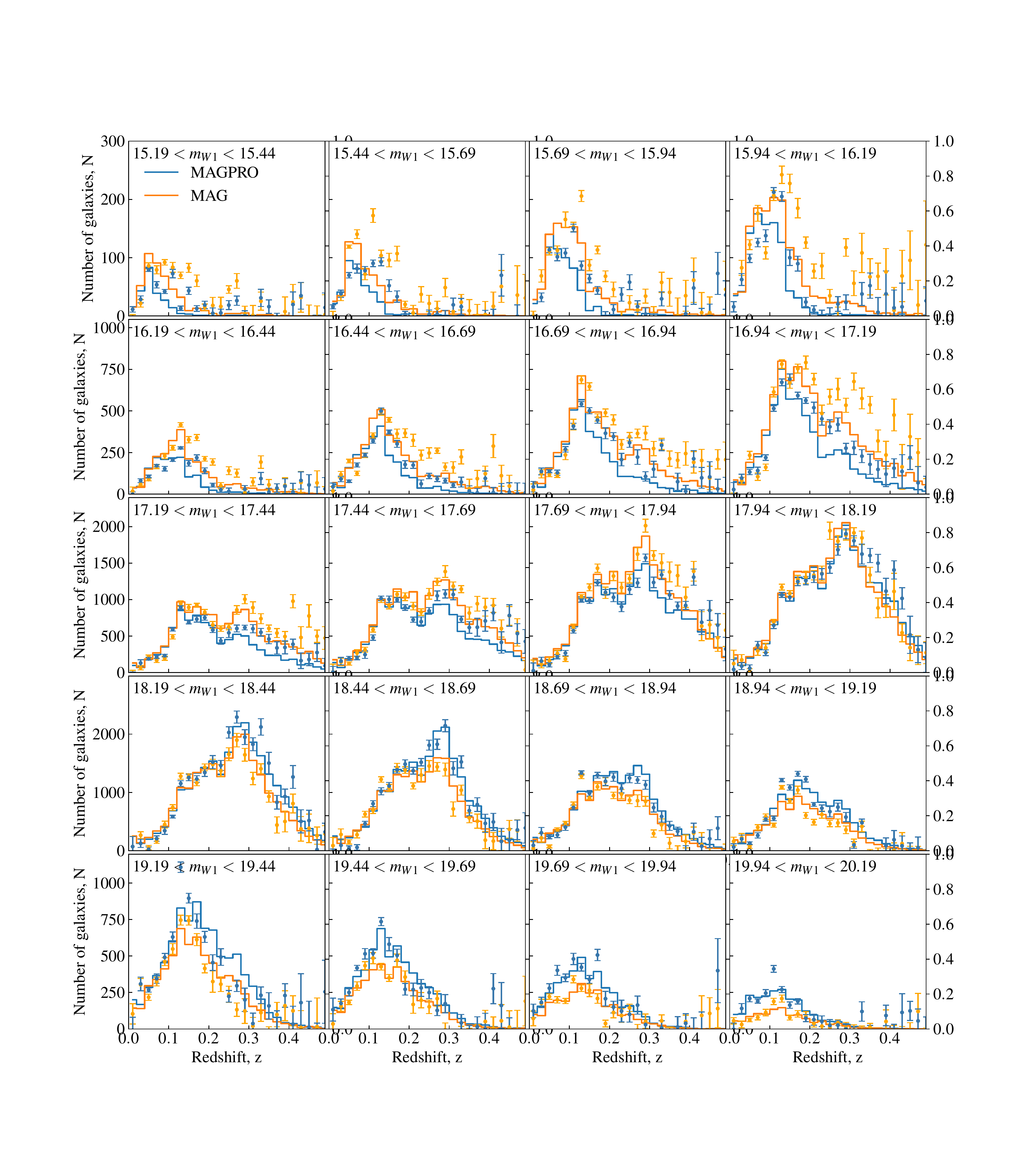}
    \caption{Redshift distribution of the GAMA galaxies binned by aperture ($MAG$) and profile fitted photometry ($MAGPRO$) are shown in orange and blue. In addition the $N(z)$ based on the A-factor normalised cluster-$z$s for each sample are shown.}
    \label{fig:PhotometryPz}
\end{figure*}

The resulting $N(z)$s, in comparison with the true redshift distribution of the $MAGPRO$ and $MAG$ samples, are shown in Fig. \ref{fig:PhotometryPz}. It can be seen how the redshift distribution from profile fitting ($MAGPRO$) results in smaller number counts than the redshift distribution from aperture fitting ($MAG$) at higher redshift for almost all magnitude bins. While the difference between the two sets is the largest at bright magnitudes, it becomes smaller with magnitude.

While in general it can be seen that the results from the cluster-$z$ analysis are in both cases able to trace the redshift distribution of their respective target sample, especially for the brighter bins the $N(z|m)$ is overestimating the true redshift distribution due to the unknown bias evolution, which appears to be independent of the sample to the first order. For fainter magnitudes the impact of the bias becomes smaller and the $N(z|m)$ is in close agreement with the true redshift distributions.

As the cluster-$z$s are in both cases able to trace the true underlying redshift distribution of the target data, with the same uncertainties and systematics, we conclude that the bias due to missed flux on the cluster-$z$s is small.

\section{Measurement of k-corrections and mass-to-light ratios}
For the transformation of the observers' frame GLF to GSMF we use k-corrections and mass-to-light ratios from \Mentari \, and \cite{Jarrett2023} which we display in the following.

\subsection{K-corrections}
K-corrections are part of the relation between observed apparent magnitude to emitted- or rest-frame absolute magnitudes \citep[][]{Wilson2002} and are calculated by $K = M - m + DM - 2.5\log(1+z)$. Here $M$ indicates the absolute magnitude, $m$ the apparent magnitude and the distance modulus of an object at certain redshift $z$, $DM = 5\log(\frac{D_L(z)}{Mpc}) + 25$, with $D_L(z)$ being the luminosity distance at redshift $z$ in units of $Mpc$.

In \Mentari \, we are able to calculate $K$ for each object as the values for $M$, $m$ and $z$ are provided from the simulation. We then bin the resulting $K$ values based on the apparent magnitude and redshift of the objects using the same binning as used for the derivation of the unWISE cluster-$z$s. Thus we are able to derive the k-correction in bins of magnitude and redshift and obtain a statistical measurement for the galaxy sample within each bin. As \Mentari \, uses a logarithmic spacing in redshift, the resulting k-corrections are interpolated using a cubic spline to translate the results into the linear redshift grid used by the cluster-$z$s.

The k-correction by \cite{Jarrett2023} is a magnitude independent fit to the "Sb" galaxy template from \cite{Brown2014}. The k-corrections of the "Sb" template have been chosen by \cite{Jarrett2023} as it provides a good first order approximation for all different galaxy types. For the fit an exponential function of the form: $K = 1 - a_0 e^{\alpha z}$ has been used, with best fitting values of $a_0=1$ and $\alpha = -2.614$. As the secondary term ($2.5\log(1+z)$) is included in the k-corrections by \cite{Jarrett2023} we accounted for this in the comparison and application to the GLF. 

\begin{figure}
    \centering
    \includegraphics[width=\columnwidth]{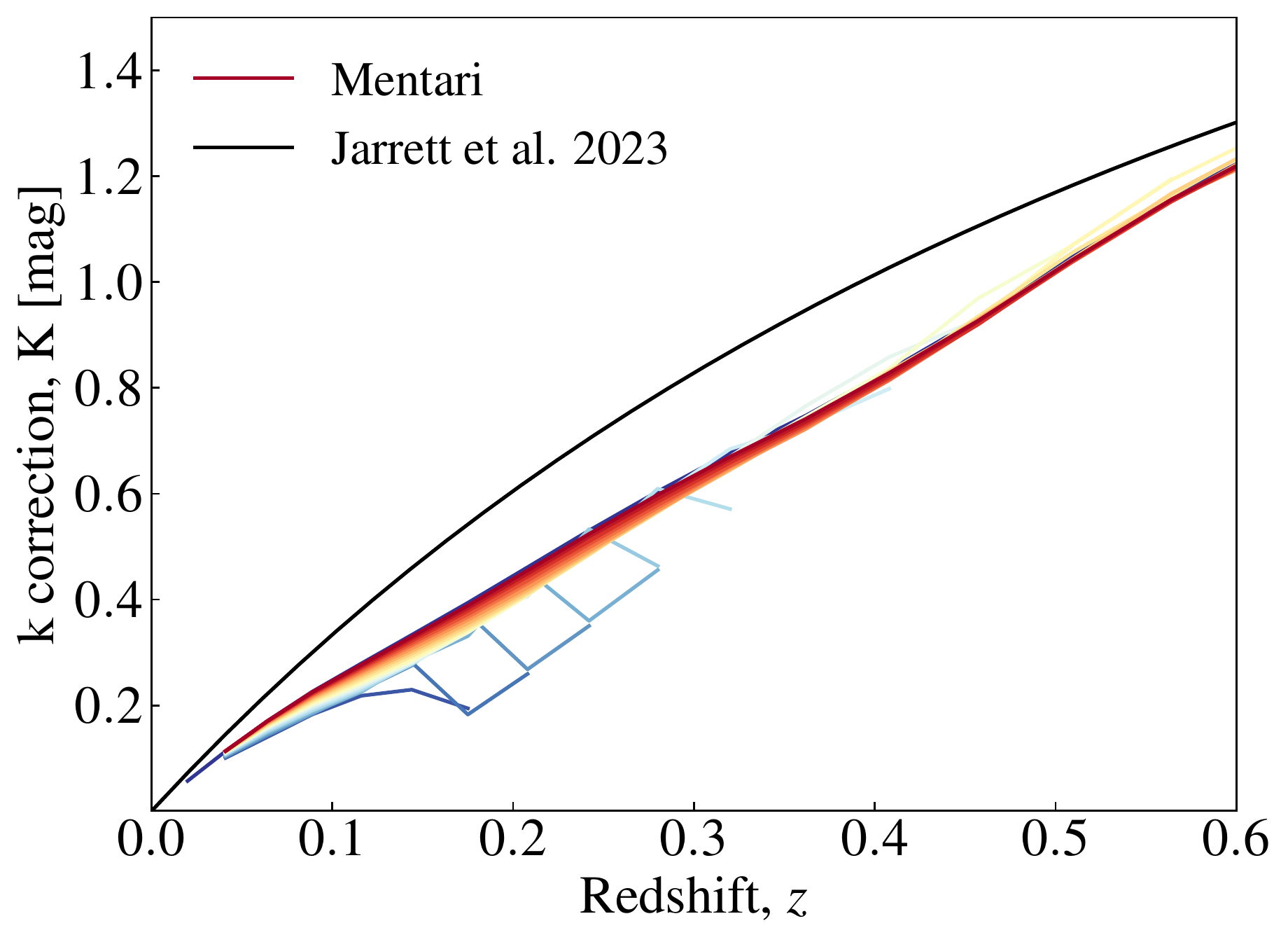}
    \caption{K-corrections for W1 based on the "Sb" SED template by \protect\cite{Jarrett2023} (black) and from \Mentari \, as a function of redshift. The k-corrections of \Mentari \, are measured in bins of magnitude indicated by colors from blue (bright) to red (faint). While the main slope between the two corrections is similar, there is a systematic offset of $\sim0.25$ mag at intermediate redshifts.}
    \label{fig:kcorr}
\end{figure}

In Fig. \ref{fig:kcorr} the resulting k-corrections are shown, and it can be seen that the slopes by \cite{Jarrett2023} is steeper than \Mentari \, with a maximum offset of $\sim 0.25 mag$ at $z\sim 0.3$. At higher redshift the k-corrections of \cite{Jarrett2023} becomes flatter and the offset becomes smaller.

At most magnitudes the k-corrections from \Mentari \, follow a similar trend, which supports the use of one galaxy template as a first order estimate of the k-corrections. The trend of showing a flatter slope for the brightest bins of \Mentari \, could be a bias due to low number statistics in \Mentari \, but also highlights the need to consider all different galaxy types for the derivation of k-corrections.

\subsection{Mass vs light}
If the emitted light of an object as well as its stellar mass is known, the mass-to-light ratio follows trivially. We calculate the luminosity $L_*$ for each object in \Mentari \, with its given $M$ using the bolometric magnitude of the sun in the W1-band of $M_{bol,\odot, AB} = 5.939$ according to $\frac{L_{W1}}{L_\odot} = 10^{0.4(M_{bol,\odot} - M)}$. As the stellar mass $M_*$ for each object is directly available from \Mentari \, the mass-to-light ratio $ML = M_*/L_{W1}$ follows directly. The mass-to-light ratio is then calculated for each redshift slice of \Mentari \, in bins of luminosity with $\Delta \log(L_{W1}) = 0.25$. As the final step, the resulting grid is interpolated using a cubic spline to obtain the same redshift and magnitude spacing as the cluster-$z$s.

The mass to light scaling relation form \citep{Jarrett2023} is described by a third order polynomial $\log(M_*) = A_0 + A_1 \log(L_{W1}) + A_2 \log(L_{W1})^2 + A_3 \log(L_{W1})^3$, with $A_i$ coefficients of $-12.62185$, $5.00155$, $-0.43857$, and $0.01593$ respectively.

It is seen that the scaling relation from \Mentari \, shows a general offset to the results from \cite{Jarrett2023}. Since the scaling relation from \cite{Jarrett2023} is based on observational data, we correct the results from \Mentari \, by a scalar of $ -0.47 \log(M_*)$, which has been derived by a simple least squares analysis between the two relations, to account for this offset.

\begin{figure}
    \centering
    \includegraphics[width=\columnwidth]{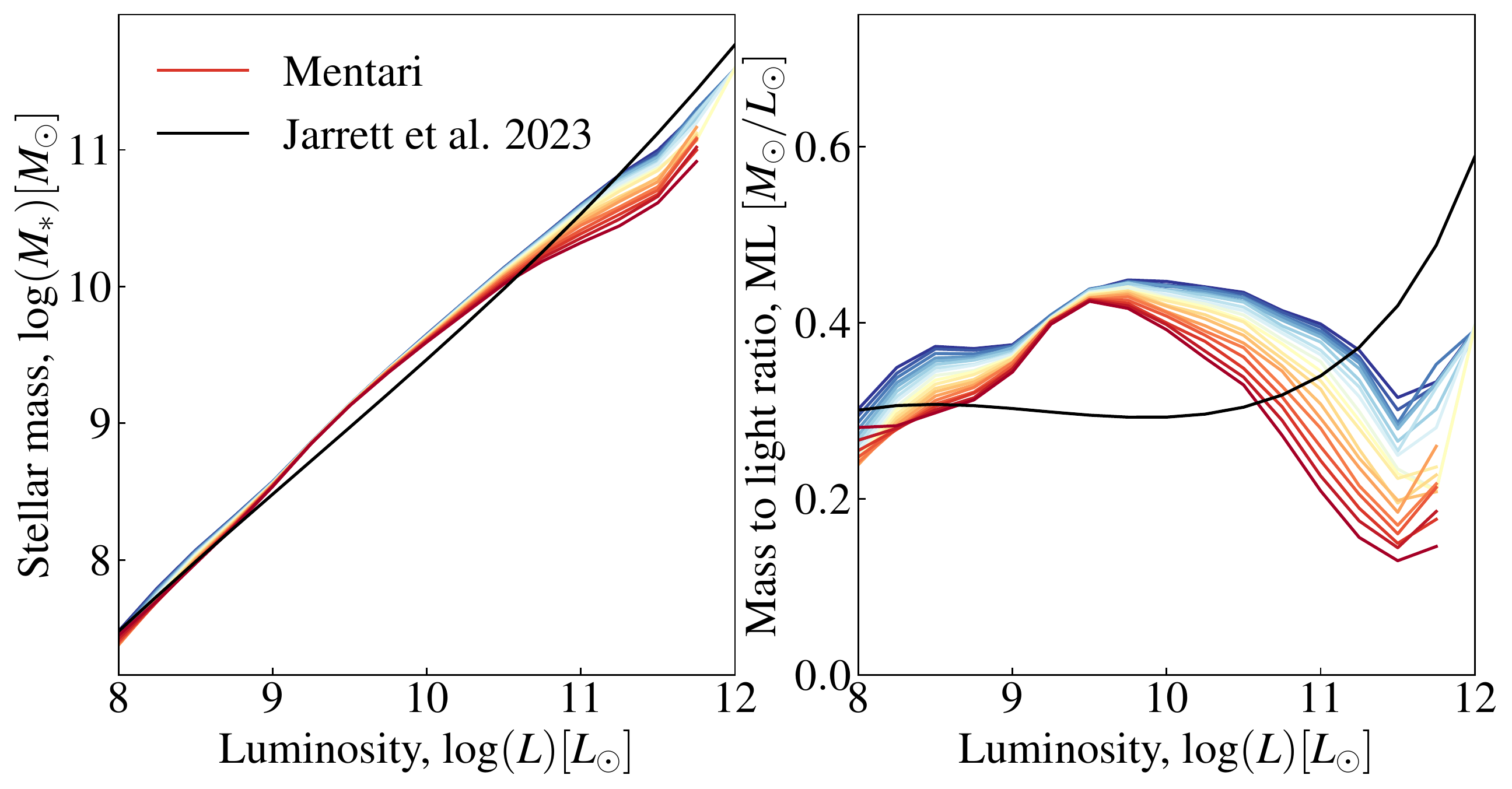}
    \caption{In the first panel the W1 scaling relation from \Mentari \, in bins of redshift are shown in comparison to results from \protect\cite{Jarrett2023}. The second panel shows the resulting mass to light ratios for the same data as the first panel. Despite the different shapes of the individual mass to light ratios, both measurements show a similar scaling relation.}
    \label{fig:MLratio}
\end{figure}

In the left panel of Fig. \ref{fig:MLratio} the corrected scaling relation from \Mentari \, and \cite{Jarrett2023} are displayed. Apart from small offsets, it can be seen that the slopes are in good agreement up to $L_{W1} \sim 10^{11} L_\odot$, where the results from \Mentari \, show a redshift-dependent downturn. 

While the scaling relations themselves are in general agreement, the individual mass to light ratio values show inverse shapes with mean differences of $\sim 0.1 M_\odot/L_\odot$ between the two measurements, highlighting the small differences within the scaling relation from the first panel.

\section{The unknown bias evolution of the target dataset}
\label{app:UnknownBias}
As discussed in Sec. \ref{sec:discussion}, the largest uncertainty within the clustering technique is the evolution of the unknown sample galaxy bias. Due to the normalisation factors $A_m$ any flat bias is of no concern to our data, but any evolution - especially with redshift - would cause an over- or underestimation of the true redshift distribution. 

To get an estimate of the unknown bias and its evolution, we calculate the auto-correlation function for data from the Theoretical Astrophysical Observatory (TAO) \citep[][]{Bernyk2016} and WISECat as a proxy for the bias.

TAO is an online virtual laboratory which contains mock observations of galaxy survey data. Using TAO we create a light-cone with a $100\mathrm{deg}^2$ field up to $z<0.7$ of galaxies with W1 magnitudes. The mock-catalogue is based on the MultiDark simulation by \cite{Klypin2016} using Planck cosmology \citep[][]{Planck2013} and the Semi-Analytic Galaxy Evolution (SAGE) model \citep[][]{Croton2006}. The spectral energy distribution is modeled using a Chabrier IMF \citep[][]{Conroy2009}.

\begin{figure}
    \centering
    \includegraphics[width=\columnwidth]{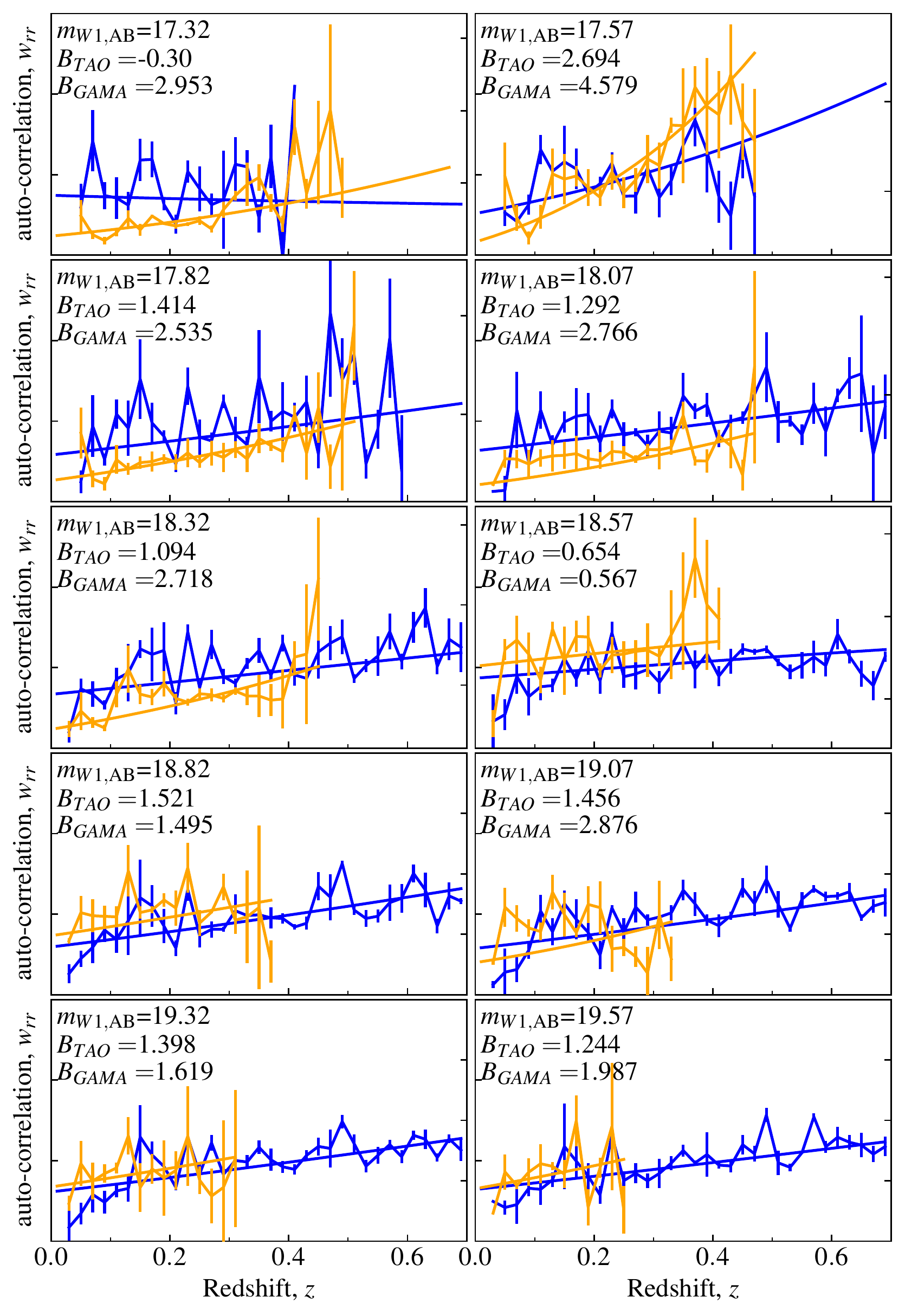}
    \caption{Auto-correlation function and bias evolution model for TAO (blue) and WISECat (orange) in bins of magnitude. The value of the evolution parameter $B$ is shown in the upper left corner of each panel for each dataset. It can be seen that the slopes are qualitatively in agreement.}
    \label{fig:BiasSlopes}
\end{figure}

We model the bias evolution by using a simple parametrisation $w_{rr} = A\times[1+z]^B$, where $A$ is a marginalisation factor and $B$ the evolution parameter. In Fig. \ref{fig:BiasSlopes} the measured auto-correlation functions $w_{rr}$ and the best fitting models are shown. Here the $w_{rr}$ of both datasets are arbitrarily scaled for comparison as the focus is on the evolution with redshift. It can be seen that the slopes, and therefore the values of the evolution parameters, are qualitatively similar and no systematic difference between the datasets, as well as no clear trend for an evolution with magnitude, can be found. For both datasets the bias evolution results in positive values of $B$, which indicates that the cluster-$z$s are overestimating the true redshift distribution with redshift. This observation is in agreement with earlier studies \citep[e.g.][]{Menard2013, Rahman2016b}. 

With $\bar{b}_u \propto (1+z)^B$ we can calculate the change of $\Phi$ according to $B$. Assuming $B \sim 1.5 - 2$, $\Phi$ would experience a change by a factor of $\sim 2 - 2.5$ at $z=0.6$ due to the evolution of the bias. This change can be included into the density evolution $P$, where it would result in a decrease of $P$ by $\sim 1.25 - 1.65$, which results in a change of $\Delta \log(\phi^*) \lesssim 0.4 $ at $z=0.6$.

We decided not to apply the bias correction to our analysis as we cannot be certain about how well TAO is able mimic the true bias evolution of the unWISE galaxies as well as how representative the WISECat galaxies are for the full unWISE distribution over the full redshift range of our study. Nevertheless, this analysis shows a path to using simulations to calibrating the impact of bias evolution.

\label{lastpage}
\end{document}